\documentclass[11pt,english]{article}
\pdfoutput=1
\usepackage{jheppub,empheq}
\usepackage{hyperref}
\usepackage{simplewick}
\usepackage{verbatim}
\usepackage{subfigure}
\usepackage{graphics, color}
\usepackage{graphicx}
\usepackage{mathabx}
\usepackage{amsmath,mathtools}
\usepackage{amssymb}
\usepackage{amsthm}
\usepackage{appendix}
\usepackage{array}
\usepackage{tabularx}
\usepackage{amsfonts}
\usepackage{lipsum}   
\usepackage[usenames,dvipsnames]{xcolor}
\usepackage[normalem]{ulem}
\usepackage{braket}
\usepackage{titlesec}

\definecolor{darkgreen}{rgb}{0,0.5,0}

\newcommand{\be}{\begin{equation}}
\newcommand{\ee}{\end{equation}}

\newcommand{\uu}{\hat{u}}

\begin{document}
\title{Holographic non-Fermi liquids at large $d$}

\author{Shamit Kachru and Milind Shyani}
\affiliation{Stanford Institute for Theoretical Physics, Stanford, CA 94305, USA}
\abstract{
Motivated by the results of Dynamical Mean Field Theory, we study the two-point function of fermions moving in a charged black brane background in $AdS_{d+1}$ in the limit of large $d$.  We 
observe the emergence of a locally critical form of the fermion self-energy, with a strongly constrained range of possible scaling behaviors at large $d$.  Novelties compared to the analysis in $d=3,4$ include an enlarged regime of temperatures where the results apply, and the analytical tractability of the calculations.
}

\maketitle

\section{Introduction}

Non-Fermi liquids are of interest in condensed matter physics 
both as abstract possible states of quantum matter, and for potential application
to phases seen in modern materials.  The arsenal of tools available to study
non-Fermi liquids is quite limited, however.  One such tool, which has been
applied extensively to this problem in the last decade, is holography.  In this
note, we discuss the possible applications of holography to the theory of fermions at finite density in $d$ space-time dimensions as $d \to \infty$.

We have two motivations for studying large $d$ holography.
One arises directly from the condensed matter literature:
the dynamical mean field theory (DMFT) ansatz works by considering fermions hopping on a lattice of coordination number $z$ in the limit of large $z$.  For traditional
lattices (e.g. the cubic lattice), this coincides with a limit of large dimension.  The DMFT ansatz is a widely used tool in condensed matter theory, and has found spectacular success in describing strongly coupled phenomena such as the metal/non-metal Mott transition \cite{Georges:1996zz,vollhardt}. The ansatz uses the assumption that the lattice self energy is spatially local. For a quasiparticle near a Fermi surface, in momentum space, this means that the Green's function,
\begin{align}
    G(\omega,k) = \frac{Z(\omega,k_F)}{\omega - v_F k_\perp + \Sigma(\omega,k)},
\end{align}
has a self-energy that only depends on the frequency,
\begin{align}
    \Sigma(\omega,k) \equiv \Sigma(\omega). \label{dmft}
\end{align}
Here the dependence on the intensive parameters of the system is left implicit. It is a proven fact that the self energy becomes spatially local at infinite lattice coordination number $z$. However, DMFT has found enormous success even when the coordination number is as small as $z=6$.

The qualitative reason for the emergence of a purely $\omega$-dependent self-energy in DMFT is the self-consistency of a mean-field like single site approximation that governs the fermion dynamics.  Particularly interesting to us are the situations -- relevant to non-Fermi liquid dynamics -- where an approximate conformal quantum mechanics emerges at intermediate energies.  Part of our
interest in studying large $d$ holography is to witness the dual phenomenon -- the emergence of a (dual) $AdS_2$ geometry at large $d$.\footnote{A distinct -- and very successful -- application of holography to a similar class of problems occurs in the study of the Sachdev-Ye-Kitaev models. For a review with further references, see \cite{Rosenhaus:2018dtp}.}
Importantly, from the DMFT analysis, we might expect that the emergent geometry is now applicable to the problem over a wider range of energy scales than has been traditionally observed in the direct analysis of holographic systems dual to finite density fermions in three or four space-time dimensions.

This brings us to our second motivation for studying large $d$ holography -- analytical tractability of the physics of the emergent IR geometry which governs the finite density holographic system. Recently, the large $d$ limit of semi-classical Einstein gravity has been extensively studied in the classical GR literature \cite{Emparan:2013moa,Emparan:2013xia,Emparan:2014aba,Emparan:2014cia,Emparan:2015rva,Bhattacharyya:2015dva,Bhattacharyya:2015fdk,Dandekar:2016fvw,Bhattacharyya:2017hpj}. The idea is that the number of spacetime dimensions $d$, provides a new parameter for a perturbative analysis of standard problems in classical GR. One such example is finding the spectrum of black hole  quasinormal modes. This is an extremely hard computation at fixed $d$, with very few results known analytically \cite{Kovtun:2005ev,Kokkotas:1999bd}. However, such computations can be done with relative ease in a large $d$ perturbation expansion. In fact, results of the quasinormal spectrum using large $d$ perturbation theory are in remarkable agreement with numerics even for spacetime dimensions $d=6$.

In sum, then, the aim of this paper is to use large $d$ perturbation theory to understand strongly coupled conformal field theories at finite charge density and temperature. We calculate the scalar and the fermion two-point function in the boundary CFT by solving the large $d$ Einstein gravity equations in the holographic bulk dual perturbatively. The full two-point function can be explicitly obtained by using matched asymptotic expansions perturbatively at large $d$. 
The tractability of the $1/d$ expansion, including computations beyond leading order at large $d$, is an advantage the present framework has over the current state-of-the-art in DMFT, where explicit $1/z$ corrections have been difficult to compute.

\subsection{Overview of results}

The holographic setup at finite charge density and temperature has been considered previously for the case of fixed $d$ in the seminal work of Faulkner et. al \cite{Faulkner:2009wj,Faulkner:2010da,Faulkner:2011tm,Faulkner:2013bna}. They studied the two-point function of scalars and fermions in a near-extremal black brane background in $AdS_{d+1}$. This corresponds to a conformal field theory on the spatial manifold $\mathbb R^{d-1}$ at finite temperature and chemical potential. It is well known that such a near-extremal black brane grows a nearly $AdS_2$ geometry in its near-horizon region.\footnote{In many situations, one would expect the $AdS_2$ throat to be valid down to a low energy cutoff, beyond which the geometry is infrared completed by onset of an instability.  Such a cutoff is naturally $1/N$ suppressed, as well as being suppressed by dynamical scales in various concrete scenarios.  We therefore work with the $AdS_2$ geometry without further apology.} 
The $AdS_2$ geometry in turn can be thought as having its own holographic $CFT_1$ dual. It was found that the two-point function in the $CFT_d$ can be written as,
\begin{align}
    G_{UV}(\omega,k) = \frac{Z(\omega,k_F)}{\omega - v_F k_\perp + G_{IR}(\omega,k_F)}, \label{mit}
\end{align}
where $G_{IR}$ is the two-point function in the $CFT_1$ that is dual to the near-horizon $AdS_2$ geometry. This is similar to the DMFT ansatz (\ref{dmft}), since the near-horizon two-point function $G_{IR}(\omega,k_F)$ serves as the local self energy $\Sigma(\omega)$ of the two-point function. Our work is morally similar to \cite{Faulkner:2009wj,Faulkner:2010da,Faulkner:2011tm,Faulkner:2013bna}, but differs in the details. 

The first difference concerns the range of applicability of our results. At large $d$, we find that the two-point function can be tuned to take the form (\ref{mit}) for any $\frac{T}{\mu} \sim O(1)$, in large $d$ counting. This is parametrically larger than the range of applicability in Faulkner et. al where $\frac{T}{\mu} \rightarrow 0$. This extended regime of validity is interesting from the DMFT perspective, since the ansatz (\ref{dmft}) is valid for any finite $\frac{T}{\mu} $.

The other difference is the ability to compute quantities explicitly by using the large $d$ perturbation theory in the bulk. For instance, one of the central results of \cite{Faulkner:2009wj,Faulkner:2010da,Faulkner:2011tm,Faulkner:2013bna} is about the form of the IR two-point function $G_{IR}(\omega,k)$. For scalars it was shown to take the form,
\begin{align}
    G_{IR} (\omega,k_F) \, \, \propto \, \, T^{2\nu},     
\end{align}
where,
\begin{align}
    \nu = \sqrt{\frac{m^2}{d(d-1)} - \frac{q^2}{2d(d-1)} + \frac{k_F^2}{d(d-1)r_h^2} + \frac{1}{4}}.
\end{align}
The parameter $\nu$ is related to several critical exponents in the theory and is of central importance.
%about a prediction in the large $d$ setup. The IR two-point function for scalars in (\ref{mit}) scales as,
%where the critical exponent $\nu$ is given by,
This exponent can be tuned to any value by varying $m$ or $q$. At large $d$, we find that if we limit ourselves to almost marginal $\left(\Delta \sim d\right)$ or relevant deformations $\left( \Delta < d\right)$ in the $CFT_d$, the scalar two-point function can be written in the form of (\ref{mit}) only if,
\begin{align}
    \nu = \frac{V(m,q,k_F)}{d^{1/3}},
\end{align}
or smaller, were $V(m,q,k_F)$ is an $O(1)$ number, with at the most a logarithmic dependence on $d$. While it is still true that $V(m,q,k_F)$ can be tuned to any $O(1)$ value by varying $m$ or $q$, the fact that $\nu$ is small in the large $d$ limit is a fact that arises on solving the bulk equations of motion. A similar result holds true for fermions.\footnote{\label{bose}For readers confused by the existence of parameters where the scalar two-point function exhibits a surface in momentum space, be comforted that the regime where this occurs is one where the charged black brane geometry is unstable, and so the result is not a stable phase of holographic quantum matter.}

%The concrete advantage of working at large spacetime dimensions is that $d$ provides an additional parameter that allows a perturbative analysis. This fact was first appreciated in classical GR and the gravitational waves literature . Although there are some obvious issues as well. The first is that at large $d$, the short distance physics that arise in quantizing general relativity get increasingly divergent. This is because since $G_N$ has mass dimension $1-d$, and thus gets increasingly irrelevant. We will take the view point that general relativity as a classical theory can be studied in its own right in any number of spacetime dimensions. Since we are working at large $N$, we believe that the leading order classical physics is well defined. 

It is important to emphasize that although the appearance of a local self energy in both DMFT and holography might seem striking, the two results (\ref{dmft}) and (\ref{mit}) apply in two very different classes of systems that are at best spiritually related to each other. The holographic setup is a doped large $N$ conformal field theory with a sparse spectrum, which clearly isn't the kind of system that is studied in the DMFT literature. However, the fact that the same phenomenon (i.e. local self energy) presents itself in two such different setups is worthy of investigation, and analysis of one system may yield qualitative insight into behaviors seen in the other.  (This is the standard justification for using holography to yield tractable toy models of many dynamical phenomena which are thought to occur -- but are difficult to understand directly -- in conventional quantum field theories.)

Another issue that we would like to address concerns the existence of the holographic dictionary in a space-time of high dimension. It follows from Nahm's classification of superconformal algebras that there are no supersymmetric conformal field theories in spacetime dimensions $d>6$ \cite{Nahm:1977tg}. %Thus, such large $d$ models of holography cannot arise from a top-down string theory construction. 
The existence of interacting conformal field theories without supersymmetry in spacetimes of high dimension remains a very interesting open question. (For constraints on such theories, see e.g. \cite{Gadde:2020nwg}.)
We will take the approach that at infinite $N$, the bulk theory becomes purely classical, and the boundary CFT becomes a theory of generalised free fields that can be studied perturbatively using a large $d$ expansion. Studying a charged black hole at infinite $N$ then corresponds to studying the boundary generalised free field theory at finite temperature and chemical potential. It is unclear what happens to this construction at subleading orders in $N$, but our aim is to study such systems systematically in a large $d$ perturbation expansion with the hope that lessons learnt here could be applied towards understanding field theories at finite temperature and charge density at fixed $d$ some day. 

The organisation of the rest of this note is as follows.
In section \ref{larged}, we elaborate the black hole geometry at large $d$, and its near-extremal limit. In section \ref{scbulk}, we evaluate the scalar two-point function by evaluating the bulk equations of motion in different patches of this geometry. In section \ref{asympsc}, we match the solutions in the overlapping regions and find that the full two-point function takes the form (\ref{mit}) for certain parametric regimes of $\nu$. The concluding section contains some brief remarks about the relationship of physics seen in this system to analyses of other systems.  Several calculations involving other parameter regimes for the scalars, and the analysis of fermions, are relegated to appendices.

%It is worthwhile to note that the black hole quasinormal modes found in \cite{Emparan:2013moa} using large $d$ perturbation theory are in remarkable agreement with numerics even when $d=5$.

%%%%%%%%%%%%%%%%%%%%%%%%%%%%%%%%%%%%%%%%%%%%%%%%%%%%%%%%%%%%%%%%%%%%%%%%%%%%
%%%%%%%%%%%%%%%%%%%%%%%%%%%%%%%%%%%%%%%%%%%%%%%%%%%%%%%%%%%%%%%%%%%%%%%%%%%%
%%%%%%%%%%%%%%%%%%%%%%%%%%%%%%%%%%%%%%%%%%%%%%%%%%%%%%%%%%%%%%%%%%%%%%%%%%%%
%%%%%%%%%%%%%%%%%%%%%%%%%%%%%%%%%%%%%%%%%%%%%%%%%%%%%%%%%%%%%%%%%%%%%%%%%%%%
%%%%%%%%%%%%%%%%%%%%%%%%%%%%%%%%%%%%%%%%%%%%%%%%%%%%%%%%%%%%%%%%%%%%%%%%%%%%
%%%%%%%%%%%%%%%%%%%%%%%%%%%%%%%%%%%%%%%%%%%%%%%%%%%%%%%%%%%%%%%%%%%%%%%%%%%%
%%%%%%%%%%%%%%%%%%%%%%%%%%%%%%%%%%%%%%%%%%%%%%%%%%%%%%%%%%%%%%%%%%%%%%%%%%%%
%%%%%%%%%%%%%%%%%%%%%%%%%%%%%%%%%%%%%%%%%%%%%%%%%%%%%%%%%%%%%%%%%%%%%%%%%%%%

\section{$AdS_{d+1}$ black holes at large $d$} \label{larged}
As mentioned in the introduction, we are interested in understanding strongly coupled field theories at finite temperature and non-zero charge density. This corresponds to studying charged black holes in Einstein gravity minimally coupled with matter in asymptotically $AdS_{d+1}$ spacetime \cite{Chamblin:1999tk}. Since the spatial manifold of our CFT is $\mathbb R^{d-1}$, we will be interested in studying black holes with planar horizons. Such black holes are also known as black branes. We do not have a specific field theory at hand, but rather an entire class of large $N$ field theories that are strongly interacting, and have a sparse spectrum of light operators \cite{Heemskerk:2009pn}. It is widely believed that all such field theories have a universal sector that is dual to Einstein gravity in asymptotically $AdS_{d+1}$ spacetime. 

In section \ref{hfm}, we study charged black brane solutions in Einstein gravity at large $d$, and specialise to near-extremal black holes in section \ref{nearext}. The large $d$ geometry is extremely simple, and the only nontrivial physics takes place in a small region near the horizon as emphasized originally by Emparan et al. \cite{Emparan:2013xia,Emparan:2013moa,Emparan:2014cia,Emparan:2014aba,Emparan:2015rva}. In terms of holographic RG \cite{Susskind:1998dq,Heemskerk:2010hk}, this means that the boundary CFT has a nontrivial IR that is governed by the near-horizon black brane geometry. We will find that at large $d$, the near-horizon geometry exists for all $O(1)$ values of $T/\mu$.

%This is similar to the work done on semi-local quantum criticality \cite{Faulkner:2009wj,Faulkner:2010da,Faulkner:2011tm,Faulkner:2013bna}, where it was found that nontrivial IR physics is governed by the near-horizon $AdS_2$ geometry. 
% Our motivation is in the range of phenomena that are governed by the near-horizon geometry. We find that the physics of all energy scales $\omega/\mu \sim O(1)$ is governed by the near-horizon geometry at large $d$. This is parametrically larger than the range of applicability of \cite{Faulkner:2009wj,Faulkner:2010da,Faulkner:2011tm,Faulkner:2013bna} where $\omega/\mu \rightarrow 0$.

\subsection{Charged black brane in \texorpdfstring{$AdS_{d+1}$}{AdS}} \label{hfm}
The bulk Einstein-Maxwell action in asymptotically $AdS_{d+1}$ spacetime is given by,
\begin{align}
    S=\frac{1}{16 \pi G_N} \int d^{d+1}x \sqrt{-g} \left( R - \frac{\ell^2}{e^2} F^2 + \frac{d(d-1)}{\ell^2}\right). \label{action}
\end{align}
The charged black brane metric is given by the Reissner-Nordstrom (RN) solution,
\begin{align}
    ds^2   = - \frac{r^2}{\ell^2} f(r)  dt^2   + \frac{\ell^2}{r^2} \frac{dr^2}{f(r)} + \frac{r^2}{\ell^2} dx^2_{d-1}, \label{RN}
\end{align}  
where $f(r)$ is given by,
\begin{align}
    f(r) \equiv & 1  - \frac{m_b}{r^{d}} + \frac{q_b^2}{r^{2d-2}}. 
\end{align}

The parameters $m_b$ and $q_b$ are related to the mass and charge of the black brane. $\ell$ is the dimensionful curvature radius of AdS, and sets the fundamental length scale in our problem. The vector potential is given by,
\begin{align*}
     A_t  =  \frac{e q_b}{\ell^2 r_h^{d-2}} \left(1- \left(\frac{r_h}{r}\right)^{d-2} \right) \sqrt{\frac{d-1}{2(d-2)}}, \qquad A_i =0 , % \qquad \mu = \frac{e q  }{\ell r_h^{d-1}} \sqrt{\frac{d-1}{2(d-2)}} % = \left[ 1-\left(\frac{r_h}{r}\right)^d \right ] - u^2 \left(\frac{r_h}{r}\right)^d\left[ 1 - \left(\frac{r_h}{r}\right)^{d-2} \right]. 
\end{align*} 
 where $e$ is a dimensionless gauge coupling. Note that our action has an overall $G_N$ that sits in front of both the Einstein-Hilbert and Maxwell terms. In spacetime dimensions greater than four, both gravity and electromagnetism are irrelevant. The ratio of the strengths of these two forces between two particles of unit charge and unit mass (measured in units of $\frac{1}{\ell}$) is set by the gauge coupling $e$. The radius of the event horizon $r_h$ is related to $m_b$ and $q_b$ via the largest zero of the emblackening factor,
\begin{align}
    f(r_h)=0, \qquad m_b = r_h^d \left(1 + \frac{q_b^2}{r_h^{2d-2}} \right).
\end{align}

Thus every charged black brane solution is labelled by two parameters, the charge $q_b$ and the radius of the event horizon $r_h$. Since black branes are thermodynamic objects in the large $N$ limit, we can interchangeably work with microcanonical and canonical ensembles. In the canonical ensemble, the solution is labelled by two intensive parameters -- the inverse temperature $\beta$ and the chemical potential $\mu$. They are given in terms of $q_b$ and $r_h$ by,
\begin{align}
    T = \frac{d r_h}{4\pi \ell^2} \left( 1 - \frac{(d-2)q_b^2}{d r_h^{2d-2}}\right), \qquad \mu = \frac{e q_b}{\ell^2 r_h^{d-2}} \sqrt{\frac{d-1}{2(d-2)}}. \label{tmu}
\end{align}
The entropy density of the black brane is given by the Bekenstein-Hawking area law,
\begin{align}
    s = \frac{S}{A}= \frac{1}{4 G_N} \left(\frac{r_h}{\ell}\right)^{d-1}.
\end{align}

We would now like to study this solution at large spacetime dimensions. At large $d$, the black brane geometry becomes extremely simple \cite{Emparan:2013xia,Emparan:2013moa,Emparan:2014cia,Emparan:2014aba,Emparan:2015rva}. To see that, let us rewrite the emblackening factor as, 
\begin{align}
     f(r) = \left[ 1-\left(\frac{r_h}{r}\right)^d \right ] - u^2 \left( \frac{d}{d-2}\right)\left(\frac{r_h}{r}\right)^d\left[ 1 - \left(\frac{r_h}{r}\right)^{d-2} \right],
\end{align}
where we have introduced the extremality parameter,
\begin{align}
u \equiv q_b r_h^{1-d} \sqrt{\frac{d-2}{d}}. 
\end{align}
\begin{figure}
    \centering
    \includegraphics[scale=0.15]{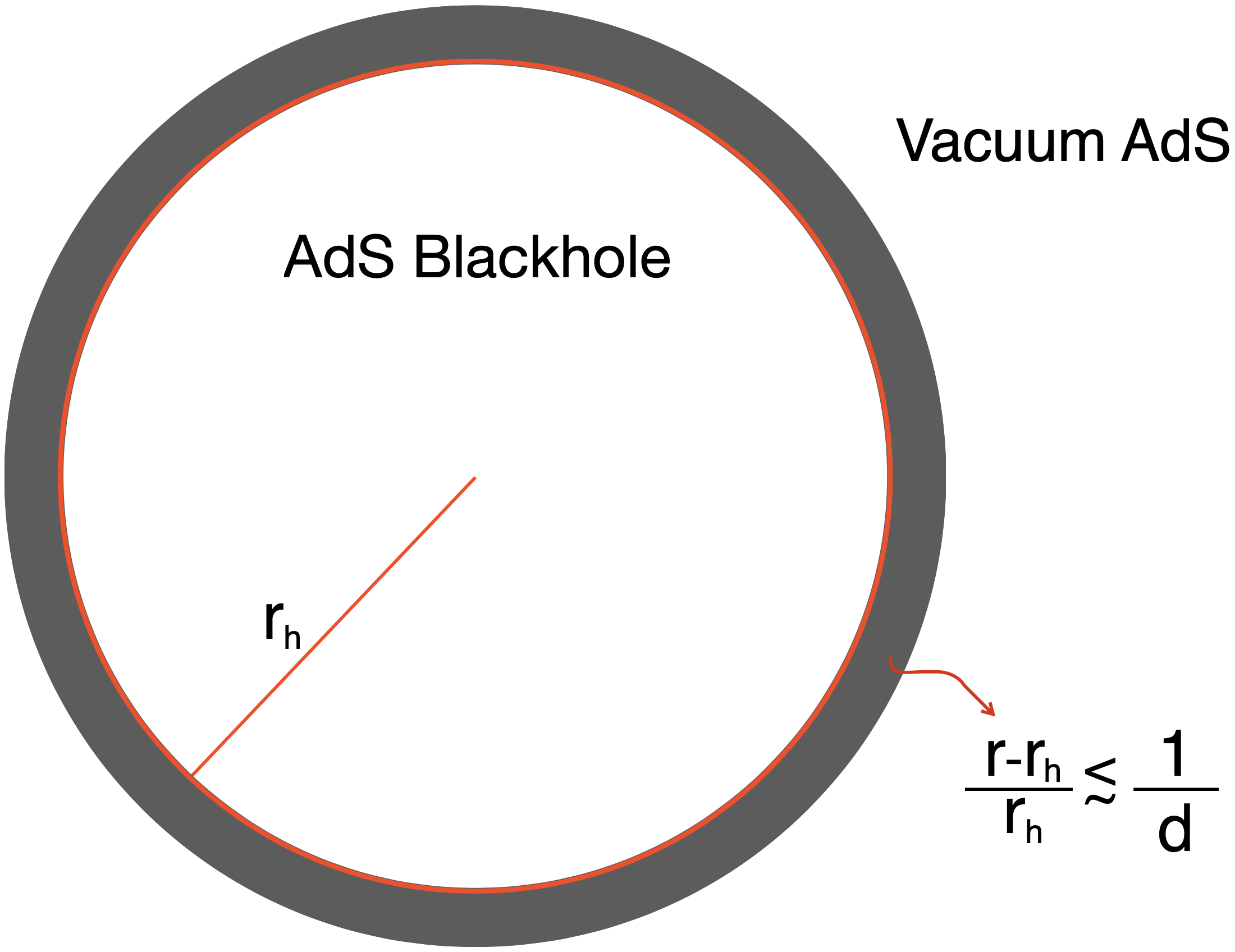}
    \caption{$f(r)$ is non-trivial in the shaded region. Outside this region, also called the sphere of influence \cite{Emparan:2013xia}, the geometry becomes that of vacuum AdS. The black brane has a similar region, but with a planar topology, which will also be called the sphere of influence.}
    \label{fig1}
\end{figure}
The black brane is neutral when $u=0$ and extremal when $u=1$. In the limit of large $d$, as we move further away from the event horizon the emblackening factor goes to unity exponentially fast in $d$. The emblackening factor is non-trivial only when,
\begin{align}
    \frac{r-r_h}{r_h} \lesssim O\left(\frac{1}{d} \right). \label{sph}
\end{align}
For all other values of $r$, such that $\frac{r-r_h}{r_h} \gtrsim O(1)$ the emblackening factor is exponentially close to one,
\begin{align}
    f(r) = 1 + O\left(e^{-d}\right).
\end{align}
Thus outside the region (\ref{sph}), the black brane geometry (\ref{RN}) reduces to that of vacuum $AdS_{d+1}$ with a constant vector potential,
\begin{align}
     ds^2   = - \frac{r^2}{\ell^2}  dt^2   + \frac{\ell^2}{r^2} dr^2 + \frac{r^2}{\ell^2} dx^2_{d-1}, \qquad A_t = \frac{e r_h u}{\sqrt 2 \ell^2}. \label{vac}
\end{align}

Inside the region (\ref{sph}), called the sphere of influence in figure \ref{fig1}, the geometry stays non-trivial. To see that we first make a coordinate redefinition,
\begin{align}
\rho \equiv \left(\frac{r}{r_h}\right)^{d-2}.
\end{align}
Taking the limit $d\rightarrow \infty$, while keeping $r_h$ fixed and $\rho \ll e^{d-2}$, the geometry in the sphere of influence becomes,
\begin{align}
ds^2_{d+1} =  -\frac{(\rho - u^2)(\rho-1)}{\rho^2} \frac{r_h^2}{\ell^2} dt^2 + \frac{\ell^2 d\rho^2}{d^2(\rho - u^2)(\rho-1)}   +  \frac{r_h^2}{\ell^2} dx_{d-1}^2, \quad A_t = \frac{e r_h u}{\sqrt 2 \ell^2}\left( \frac{\rho -1}{\rho} \right), \label{drn}
\end{align}
We find that the near-horizon region factorises into $\mathcal M_2 \times \mathbb R_{d-1}$, where $\mathcal M_2$  is a two dimensional manifold. The metric (\ref{drn}) and (\ref{vac}) spans the entire geometry.

Before we proceed with this large $d$ geometry, we would like to remark why just taking $d \rightarrow \infty $ is problematic for us. The temperature (\ref{tmu}) of the black brane is given by,
\begin{align}
    \frac{T}{\mu} = \frac{d \left(1-u^2\right)}{2 \sqrt{2} \pi  e u},
\end{align}
which blows up linearly with $d$. This is clearly an issue if we wish to interpret this system holographically. However, if we tune the extremality parameter $u$, we can make the temperature finite. This is the topic of the following section.

%%%%%%%%%%%%%%%%%%%%%%%%%%%%%%%%%%%%%%%%%%%%%%%%%%%%%%%%%%%%%%%%%%%%%%%%%%%%%%%%%%%%%%%%%%%%%%%%
%%%%%%%%%%%%%%%%%%%%%%%%%%%%%%%%%%%%%%%%%%%%%%%%%%%%%%%%%%%%%%%%%%%%%%%%%%%%%%%%%%%%%%%%%%%%%%%%
%%%%%%%%%%%%%%%%%%%%%%%%%%%%%%%%%%%%%%%%%%%%%%%%%%%%%%%%%%%%%%%%%%%%%%%%%%%%%%%%%%%%%%%%%%%%%%%%
%%%%%%%%%%%%%%%%%%%%%%%%%%%%%%%%%%%%%%%%%%%%%%%%%%%%%%%%%%%%%%%%%%%%%%%%%%%%%%%%%%%%%%%%%%%%%%%%
%%%%%%%%%%%%%%%%%%%%%%%%%%%%%%%%%%%%%%%%%%%%%%%%%%%%%%%%%%%%%%%%%%%%%%%%%%%%%%%%%%%%%%%%%%%%%%%%

\subsection{Near-extremal black branes at large $d$} \label{nearext}
We will now tune the extremality parameter to make the temperature finite. Let,
\begin{align}
    u= \sqrt{1- \frac{\hat u}{d}}, \label{extval}
\end{align}
where $\uu$ is an $O(1)$ number that is bounded below by zero. The temperature and the chemical potential of the black brane then becomes,
\begin{align}
    T^t= \frac{ \uu \mu }{2 \sqrt 2 e \pi }  + O\left(\frac 1 d\right) , \qquad  
    \mu = \frac{e r_h}{\sqrt 2 \ell^2} + O\left(\frac{1}{d}\right).  \label{tmu2}
\end{align}
We will use the superscript $t$ to denote that $T^t$ is the temperature measured by an asymptotic observer with worldline $\left(\frac{\partial}{\partial t}\right)^\mu$. As before, the temperature $T^t$ and $\mu$ are external parameters in the problem, and can be tuned to any value. We are interested in black branes that develop an $AdS_2$ region in the near-horizon geometry. At finite $d$, this corresponds to taking $T^t/\mu \rightarrow 0$. Such black branes are called near-extremal since this limit corresponds to taking the configuration with the largest charge for a given fixed mass. In the large $d$ limit, we find below that an $AdS_2$ region develops even when $T^t/\mu \sim O(1)$. %Said more simply, the large $d$ provides us with a dimensionless new parameter.  

\begin{figure}
    \centering
    \includegraphics[scale=0.8]{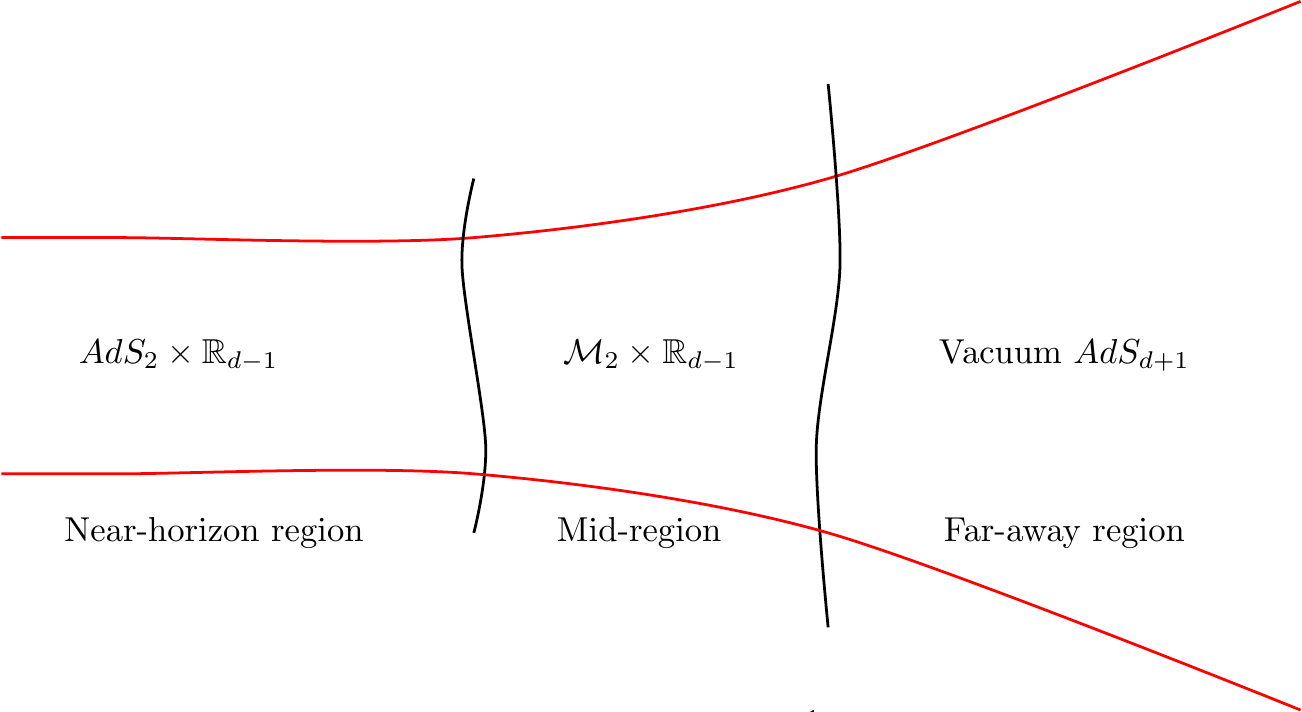}
    \caption{The black brane geometry at finite $T/\mu$ and large $d$. The mid-region geometry interpolates between the $AdS_2 \times \mathbb R_{d-1}$ throat in the deep interior and vacuum $AdS_{d+1}$ in the UV.}
    \label{adsfig}
\end{figure}

Far away from the horizon the geometry becomes vacuum AdS, but the near-horizon region becomes more interesting. The near-horizon geometry (\ref{drn}) $\mathcal M_2 \times \mathbb R_{d-1}$ mentioned in the previous section, itself develops an $AdS_2$ throat in its deep interior as depicted in figure \ref{adsfig}. We have thus three obvious regions of interest in our geometry. The far-away region which is vacuum $AdS_{d+1}$, the middle region $\mathcal M_2 \times \mathbb R_{d-1}$ and the near-horizon $AdS_2 \times \mathbb R_{d-1}$. The coordinates used in these three different regions can be found in table \ref{tabco}.

To access the $AdS_2$ region we will first use an intermediate set of coordinates $(\tilde z, \tilde \tau)$, that are defined by,
\begin{align}
    \frac{1}{\tilde{z}} \equiv \left(\frac{r}{r_h}\right)^{d-2} - u^2 = \rho - 1 + \frac{\hat u }{d},  \qquad  \tilde{\tau} \equiv t  d. \label{coord}
\end{align}
The limit $ \tilde z \rightarrow 0$ corresponds to going towards the middle region, while $\tilde z_h = \frac{d}{\hat u}$ corresponds to the event horizon. In these coordinates the metric (\ref{drn}) becomes,
\begin{align}
   ds^2 = \frac{1}{d^2} \left( - \frac{(1- \frac{\uu \tilde z}{d} )}{\left( 1 + \tilde z - \frac{\uu \tilde z}{d} \right)^2} \frac{r_h^2}{\ell^2} d \tilde \tau^2 + \frac{\ell^2 d\tilde z^2}{\tilde z^2\left(1- \frac{\uu \tilde z}{d} \right)}    \right)  + \frac{r_h^2}{\ell^2} dx_{d-1}^2. \label{nh}
\end{align}
% Since the horizon of the black hole is when $z \sim O(d)$, 
We will now make a final coordinate change to obtain the near-horizon $AdS_2$ geometry, 
\begin{align}
    z \equiv \tilde z \lambda, \quad  \tau \equiv \lambda \tilde \tau, \label{tempco}
\end{align}
where $\lambda$ is an arbitrary parameter. Finally taking the limit $\lambda \rightarrow 0$, while keeping $d \lambda \equiv \kappa$ fixed, we obtain,
\begin{align}
   ds^2 = \frac{1}{d^2} \left( - \frac{\left(1- \frac{z}{z_h} \right)}{z^2} \frac{r_h^2}{\ell^2} d\tau^2 + \frac{\ell^2 dz^2}{z^2\left(1- \frac{z}{z_h}\right)}    \right)  + \frac{r_h^2}{\ell^2} dx_{d-1}^2, \label{nh1}
\end{align}
where, 
\begin{align}
    z_h = \frac{\kappa}{\uu}. \label{zh}
\end{align}

We find that the near-horizon geometry (\ref{nh1}) is that of a black hole in an asymptotically $AdS_2 \times \mathbb R^{d-1}$ spacetime. As mentioned previously, such a near-horizon $AdS_2$ geometry is a feature of near-extremal black holes even at finite $d$.
\begin{table}[]
    \centering
   \begin{tabularx}{0.6\textwidth} { | >{\centering\arraybackslash}X 
  | >{\centering\arraybackslash}X | }
\hline
Geometry & Coordinates    \\ 
\hline
Vacuum $AdS_{d+1}$ & $(r,t,x_i)$    \\ 
\hline
$\mathcal M_2 \times \mathbb R_{d-1}$ & $(\rho,t,x_i)$   \\ 
\hline
$AdS_2 \times \mathbb R_{d-1} $  & $(z,\tau,x_i)$   \\ 
\hline
\end{tabularx}
    \caption{The three relevant regions of our geometry and the corresponding co-ordinate systems.}
    \label{tabco}
\end{table}
The important difference is about the regime of validity at large spacetime dimensions. The temperature of the $AdS_2$ black hole is given by,
\begin{align}
    T^{\tau} = \frac{r_h}{4\pi z_h \ell^2},
\end{align}
and is an $O(1)$ fraction of the chemical potential $\mu$ (\ref{tmu2}). The temperature as measured by the observer at asymptotic infinity (\ref{tmu2}) in terms of the temperature of the $AdS_2$ black hole is given by, 
\begin{align}
    T^t = \frac{r_h d \lambda}{4\pi z_h \ell^2}  =T^\tau \times d \lambda. \label{temp}
\end{align}
We find that if $d$ were held fixed, as in Faulkner et al. \cite{Faulkner:2009wj,Faulkner:2010da,Faulkner:2011tm,Faulkner:2013bna}, the geometry (\ref{nh1}) would be applicable only when, 
\begin{align}
     \frac{T^t}{\mu} \sim \lambda \rightarrow 0.
    %  \frac{\omega_t}{\mu},
\end{align}
%where $\omega_t$ is the frequency conjugate to $t$ and $T^t$ is the temperature measured with respect $t$. Since $t$ is the asymptotic time coordinate, $\omega_t$ and $T^t$ are the quantities as measured by the boundary observer. 

The first advantage of working at large $d$ is that the regime of validity of the near-horizon geometry (\ref{nh1}) is parametrically enhanced to,
\begin{align}
     \frac{T^t}{\mu} \sim d\lambda \sim O(1). \label{extv}
     %\frac{\omega_t}{\mu},
\end{align}
The second advantage is that we have a better control over the geometry. In particular we know the exact form of the interpolating geometry between the near-horizon region and the far-away region. This allows us to perturbatively solve the bulk wave equations as shown in the next section. In the rest of the note, we will drop the superscript in the temperature $T^t$.

%%%%%%%%%%%%%%%%%%%%%%%%%%%%%%%%%%%%%%%%%%%%%%%%%%%%%%%%%%%%%%%%%%%%%%%%%%%%%%%%%%%%%%%%%%%%%%%%%%%%%%%%%
%%%%%%%%%%%%%%%%%%%%%%%%%%%%%%%%%%%%%%%%%%%%%%%%%%%%%%%%%%%%%%%%%%%%%%%%%%%%%%%%%%%%%%%%%%%%%%%%%%%%%%%%%
%%%%%%%%%%%%%%%%%%%%%%%%%%%%%%%%%%%%%%%%%%%%%%%%%%%%%%%%%%%%%%%%%%%%%%%%%%%%%%%%%%%%%%%%%%%%%%%%%%%%%%%%%
%%%%%%%%%%%%%%%%%%%%%%%%%%%%%%%%%%%%%%%%%%%%%%%%%%%%%%%%%%%%%%%%%%%%%%%%%%%%%%%%%%%%%%%%%%%%%%%%%%%%%%%%%
%%%%%%%%%%%%%%%%%%%%%%%%%%%%%%%%%%%%%%%%%%%%%%%%%%%%%%%%%%%%%%%%%%%%%%%%%%%%%%%%%%%%%%%%%%%%%%%%%%%%%%%%%%%%%%%%%%%%%%%%%%%%%%%%%%%%%%%%%%%%%%%%%%%%%%%%%%%%%%%%%%%%%%%%%%%%%%%%%%%%%%%%%%%%%%%%%%%%%%%%%%%%%%%%%%

\section{Two-point function of scalars} \label{scbulk}
Our aim in the rest of this note is to find the two-point function of scalars and fermions in the boundary $CFT_d$ using large $d$ perturbation theory, at finite temperature and chemical potential. This requires us to solve the bulk equations of motion for minimally coupled scalars and fermions in the $AdS_{d+1}$ geometry. We will exclusively work with scalars in the main text, and deal with fermions in appendix \ref{fermat}.\footnote{\label{nonrel} We note that we will consider spinors with a fixed number of complex components, and not $2^{\left[ \frac{d+1}{2} \right]}$ complex components.  This explicitly breaks Lorentz invariance. This is appropriate for the application of interest, since in the low-energy limit relevant to DMFT, spin can be considered as a global symmetry that decouples from space-time symmetries. So just as in DMFT one considers a fermion with fixed number of components as $d \to \infty$, we will consider a two component Dirac spinor that transforms under a global $SO(2,1)$ in the bulk, which is dual to a two component Weyl fermion on the boundary. } The physics of the fermions and scalars is qualitatively similar. The only difference between them comes from the kind of large $d$ scaling of $m,q,k$ that we deem to be natural for the two, as elaborated near (\ref{nusc}) and in section \ref{param}.  

Consider the 
%We consider a charged bulk scalar field in the black hole background with the corresponding temperature and chemical potential. 
%In particular we are interested in the case (\ref{tmu2}) where to leading order in $d$,
% \begin{align}
%     \frac{T^t}{\mu} = \frac{\uu}{2\sqrt 2 e \pi}.
% \end{align}
% On the boundary this corresponds to working in the single trace sector of a large $N$ Conformal field theory. The scalar field can be viewed as a generalised free field in this large $N$ field theory, and has an $O(1)$ back reaction in an $O(N)$ system. We shall thus refer to these 
%We will exclusively work with scalars in the main text, and deal with fermions in appendix \ref{appfer}. They both give rise to almost identical physics. This is %Since these probe particles form an  since after all it is the black hole geometry that governs the physics. In terms of the boundary CFT, this corresponds to the fact that it is the large $N$ part of the system, and not the $O(1)$ probe particles, that governs the physics. 
minimally coupled scalar Klein-Gordon action,
\begin{align}
    S= - \int d^{d+1}x \sqrt{-g} \left( D_\mu \phi^* D^\mu \phi + m^2 \phi^* \phi \right), \label{kg}
\end{align}
where $D_\mu \equiv \partial_\mu - i q A_\mu$ is the covariant derivative, and $m$ and $q$ are the mass and charge of the scalar field respectively. Note the absence of $\frac{1}{G_N}$ in our scalar action (\ref{kg}). This corresponds to the fact that we are considering probe excitations in the black brane geometry, and any back reaction due to the probe field is thus suppressed by $G_N$ in appropriate units. The equations of motion can be easily found by first going to Fourier space in the boundary coordinates,
\begin{align}
    \phi(r,x^\mu) = \int \frac{d^dk}{(2\pi)^d} \phi(r,k_\mu) e^{i k_\mu x^\mu}.
\end{align}
The equation of motion can be easily found using the variational principle. In the black brane background (\ref{RN}) it becomes,
\begin{align}
    -\frac{1}{\sqrt{-g}}\partial_r \left( \frac{\sqrt{-g} r^2}{\ell^2} f(r) \partial_r \phi \right) + \left( \frac{\ell^2}{r^2} k^2 + \frac{\ell^2}{r^2 f(r)} \left( \omega + \mu q \left( 1 - \frac{r_h^{d-2}}{r^{d-2}}\right) \right)^2  + m^2 \right) \phi =0. \label{sceq}
\end{align}
It is almost impossible to solve this equation of motion exactly. We will try to solve it perturbatively using matched asymptotic expansions. The idea is to first solve the bulk equation of motion in the far-region $AdS_{d+1}$, and the near-horizon region $AdS_2 \times \mathbb R_{d-1}$ separately. We then match the two solutions in the middle $\mathcal M_2 \times \mathbb R_{d-1}$ geometry, to obtain the full solution. In the following sections, we will show how this can be achieved order by order in a $1/d$ expansion. 

Although, before we proceed with that calculation we would like to elaborate on the qualitative structure of the bulk solution using the following argument. 
\begin{figure}
    \centering
    \includegraphics[scale=0.37]{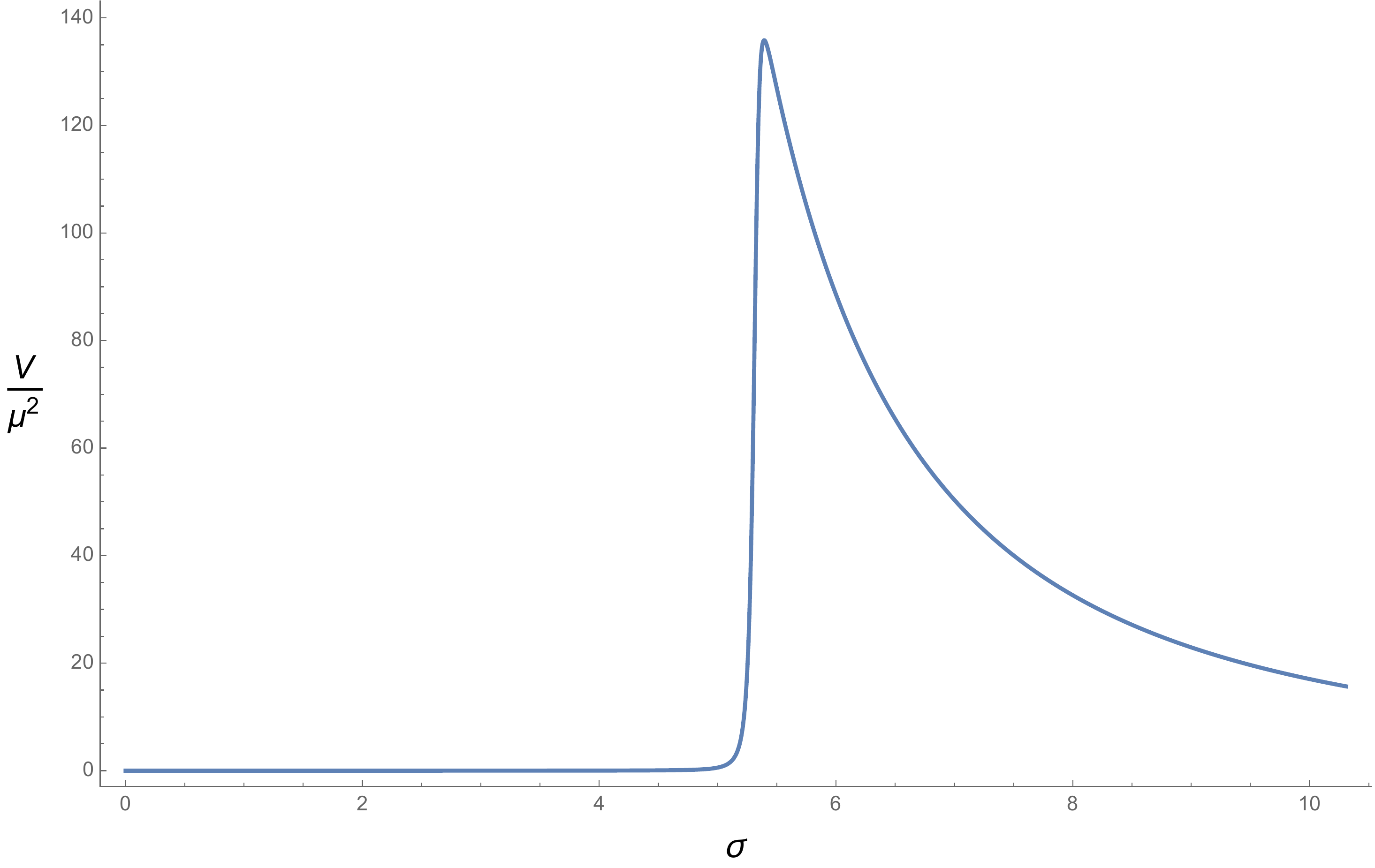}
    \caption{The scalar potential in the Schrodinger problem for the redefined field $\Phi(\sigma)$ . We have set $d=100,\ell=1,r_h=3$, and $\omega,m,k,q,\uu$ to some $O(1)$ values. $\sigma \rightarrow \infty$ corresponds to going towards the asymptotic $AdS_{d+1}$ boundary.}
    \label{scpot}
\end{figure}
The bulk Klein-Gordon equation (\ref{sceq}) can be rewritten as a Schrodinger equation by making the field redefinition, 
\begin{align}
    \phi(r,k^\mu) \equiv r^{-\frac{d+1}{2}}\Phi(r),
\end{align}
and introducing a new radial coordinate, 
\begin{align}
    \sigma(r) \equiv \int_{r_h}^r f(\tilde r) d\tilde r. \label{redef}
\end{align}
The equation of motion for the redefined field becomes,
\begin{align}
    \frac{d^2 \Phi(\sigma)}{d\sigma^2} - \left(V(\sigma) - \omega^2 \right)\Phi =0. \label{scr}
\end{align}
Plotting the scalar potential $V(\sigma)$, we find that the potential has a very sharp peak near the horizon followed by a $\frac{1}{r^2}$ fall off, as shown in figure \ref{scpot}. The height of the peak is $O(d^2)$ and it occurs near,
\begin{align}
    \frac{r - r_h}{r_h} \sim \frac{\log d}{d}.
\end{align}
Thus modes with frequencies $\frac{\omega}{\mu} \sim 1$ stay confined in the region close to the horizon, and are decoupled from the physics that occur elsewhere. This is why at large spacetime dimensions $d$, the near-horizon physics (\ref{nh1}) found in the previous section has the extended regime of validity mentioned near (\ref{extv}). The frequencies $\frac{\omega}{\mu} \sim d$ on the other hand are not confined to the near-horizon region and depend on the entire geometry. For near-extremal black branes at finite $d$, the low energy modes $\omega/\mu \rightarrow 0$ decouple from the rest of the geometry due to the fact that they are localised in the long $AdS_2$ throat that forms in the near-horizon region. The fact that $\omega/\mu \sim 1$ are also localised in the near-horizon region at large $d$, is because of the potential barrier whose height grows with $d$.

%%%%%%%%%%%%%%%%%%%%%%%%%%%%%%%%%%%%%%%%%%%%%%%%%%%%%%%%%%%%%%%%%%%%%%%%%%%%%%%%%%%%%%%%%%%%%%%%%%%%%%%%%
%%%%%%%%%%%%%%%%%%%%%%%%%%%%%%%%%%%%%%%%%%%%%%%%%%%%%%%%%%%%%%%%%%%%%%%%%%%%%%%%%%%%%%%%%%%%%%%%%%%%%%%%%
%%%%%%%%%%%%%%%%%%%%%%%%%%%%%%%%%%%%%%%%%%%%%%%%%%%%%%%%%%%%%%%%%%%%%%%%%%%%%%%%%%%%%%%%%%%%%%%%%%%%%%%%%
%%%%%%%%%%%%%%%%%%%%%%%%%%%%%%%%%%%%%%%%%%%%%%%%%%%%%%%%%%%%%%%%%%%%%%%%%%%%%%%%%%%%%%%%%%%%%%%%%%%%%%%%%
%%%%%%%%%%%%%%%%%%%%%%%%%%%%%%%%%%%%%%%%%%%%%%%%%%%%%%%%%%%%%%%%%%%%%%%%%%%%%%%%%%%%%%%%%%%%%%%%%%%%%%%%%%%%%%%%%%%%%%%%%%%%%%%%%%%%%%%%%%%%%%%%%%%%%%%%%%%%%%%%%%%%%%%%%%%%%%%%%%%%%%%%%%%%%%%%%%%%%%%%%%%%%%%%%%

\subsection{Far-away  solution}
We will first solve the bulk equation of motion (\ref{sceq}) in the far-away region. Recall that the region where the radial coordinate $r$ satisfies,
\begin{align}
    \frac{r-r_h}{r_h} \gtrsim 1 \implies \left(\frac{r}{r_h}\right)^d \gg 1,
\end{align}
is the far-away region in our terminology. As mentioned near (\ref{vac}), the geometry is just vacuum $AdS_{d+1}$ in this region. The bulk equation of motion (\ref{sceq}) takes a standard form, and the solution is given in terms of Bessel functions,

{\footnotesize
\begin{align}
    \phi(k,\omega,r) = & \left( \frac{k^2- \frac{q^2 r_h^2}{2}- \sqrt{2} q r_h \omega - \omega ^2}{4 r^2}\right)^{d/4} \left(c_1 i^{d-\Delta} \Gamma \left(\frac{d}{2}-\Delta +1\right) I_{\frac d 2 - \Delta}  \left(\frac{\sqrt{k^2-\omega ^2-\sqrt{2} q r_h \omega -\frac{q^2 r_h^2}{2}}}{r}\right) \right.   \nonumber \\
    & + \left.  c_2 i^{ \Delta } \Gamma \left(-\frac{d}{2}+\Delta +1\right) I_{\Delta -\frac{d}{2}}\left(\frac{\sqrt{k^2-\omega ^2-\sqrt{2} q r_h \omega -\frac{q^2 r_h^2}{2}}}{r}\right)\right), \label{farphi}
\end{align}}
where we have set $\ell=1$ and have defined, 
\begin{align}
    \Delta \equiv \frac{d}{2} + \sqrt{\frac{d^2}{4}+m^2}. \label{scdim}
\end{align}

The CFT two-point function is obtained by taking the ratio of the non-normalizable mode to the normalizable mode at the asymptotic boundary $r\rightarrow\infty$. Expanding the bulk solution (\ref{farphi}) as $r \rightarrow \infty$, we obtain,
\begin{align}
  \phi(k,\omega,r) =  & r^{\Delta -d} \left(i^{d-\Delta } 2^{-\frac{3}{2} (d-\Delta )} \left(2 k^2-q^2 r_h^2-2 \omega ^2-2 \sqrt{2} q r_h \omega \right)^{\frac{d-\Delta }{2}} c_1(\omega,k)+ \ldots \right)+ \nonumber  \\
  & r^{-\Delta } \left(i^{\Delta } 2^{-3 \Delta /2} \left(2 k^2-q^2 r_h^2-2 \omega ^2-2 \sqrt{2} q r_h \omega \right)^{\Delta /2} c_2(\omega,k)+\ldots \right).
\end{align}
We will be working in standard quantization and thus the mode $r^{\Delta -d}$ is the non-normalizable mode while $r^{-\Delta}$ is the normalizable mode. The two-point function is given by,
\begin{align}
    G_{UV}(\omega,k) =   2^{\frac{3}{2} (d-2 \Delta )} \left(-2 k^2+q^2 r_h^2+2 \sqrt{2} q r_h \omega +2 \omega ^2\right)^{\Delta -\frac{d}{2}} \frac{c_2(\omega,k)}{c_1(\omega,k)}. \label{farsol}
\end{align}
This is the full two-point function in a holographic $CFT_d$ at large $d$ at finite temperature and chemical potential. But of course it is still necessary to determine the values of $c_2(\omega,k)$ and $c_1(\omega,k)$. These constants are obtained by imposing infalling boundary conditions in the deep IR at $r=r_h$. This is how the solution gets to \textit{know} the IR physics. Since our solution (\ref{farphi}) is valid only in the far-away region, we need to first solve the bulk equation of motion in the near-horizon geometry and match, as we do in the following sections.

Note that, had our geometry been entirely vacuum $AdS_{d+1}$, the solution (\ref{farphi}) would be valid for all values of $r$. The two-point function in the dual vacuum $CFT_d$ is given by (\ref{farsol}), but the ratio of $c_i(\omega,k)$ is now fixed by demanding regularity in the deep interior as $r \rightarrow 0$. For completeness we write down the explicit form for the vacuum two-point function,
\begin{align}
    G_{vac}(\omega,k) = -2^{\frac{3}{2} (d-2 \Delta )} \left(2 k^2-q^2 r_h^2-2 \sqrt{2} q r_h \omega -2 \omega ^2\right)^{\Delta -\frac{d}{2}} \frac{ \Gamma \left(1-\frac{1}{2} \sqrt{d^2+4 m^2}\right)}{\Gamma \left(1+\frac{1}{2} \sqrt{d^2+4 m^2}\right)}. \label{vacgreen}
\end{align}

%%%%%%%%%%%%%%%%%%%%%%%%%%%%%%%%%%%%%%%%%%%%%%%%%%%%%%%%%%%%%%%%%%%%%%%%%%%%%%%%%%%%%%%%%%%%%%%%%%%%%%%%%
%%%%%%%%%%%%%%%%%%%%%%%%%%%%%%%%%%%%%%%%%%%%%%%%%%%%%%%%%%%%%%%%%%%%%%%%%%%%%%%%%%%%%%%%%%%%%%%%%%%%%%%%%
%%%%%%%%%%%%%%%%%%%%%%%%%%%%%%%%%%%%%%%%%%%%%%%%%%%%%%%%%%%%%%%%%%%%%%%%%%%%%%%%%%%%%%%%%%%%%%%%%%%%%%%%%
%%%%%%%%%%%%%%%%%%%%%%%%%%%%%%%%%%%%%%%%%%%%%%%%%%%%%%%%%%%%%%%%%%%%%%%%%%%%%%%%%%%%%%%%%%%%%%%%%%%%%%%%%
%%%%%%%%%%%%%%%%%%%%%%%%%%%%%%%%%%%%%%%%%%%%%%%%%%%%%%%%%%%%%%%%%%%%%%%%%%%%%%%%%%%%%%%%%%%%%%%%%%%%%%%%%%%%%%%%%%%%%%%%%%%%%%%%%%%%%%%%%%%%%%%%%%%%%%%%%%%%%%%%%%%%%%%%%%%%%%%%%%%%%%%%%%%%%%%%%%%%%%%%%%%%%%%%%%

\subsection{Near-horizon solution}
In this section we solve the bulk equation of motion (\ref{sceq}) in the near-horizon $AdS_2 \times \mathbb R_{d-1}$ geometry. This region corresponds to the case where the radial coordinate $r$ satisfies,
\begin{align}
    \left(\frac{r}{r_h}\right)^d  -  1 \lesssim \frac{1}{d}.
\end{align}
In terms of the coordinate $z$, this corresponds to the case when $z \sim O(1)$. The bulk equation of motion can be easily solved in the $AdS_2$ black hole geometry (\ref{nh1}) to obtain,

{\small
\begin{align}
    \phi(z,\omega,k) = & z^{\frac{1}{2}+\nu}   (z-z_h)^{\frac{i \omega  z_h}{r_h}}  \left(b_2(\omega,k) z_h^{-2 \nu } \, _2F_1\left(\frac 1 2  + \nu  - \frac{i q}{\sqrt 2 d},\frac 1 2  + \nu +\frac{i q}{\sqrt 2 d}  +\frac{ i  \omega }{2 \pi T^t};1+2 \nu ;\frac{z}{z_h}\right) \right .  \nonumber \\
    & \left. +  b_1(\omega,k) z^{-2\nu} \, _2F_1\left(\frac 1 2  - \nu  - \frac{i q}{\sqrt 2 d},\frac 1 2  - \nu +\frac{i q}{\sqrt 2 d}  +\frac{ i  \omega }{2 \pi T^t};1-2 \nu ;\frac{z}{z_h}\right)\right). \label{nearsol}
\end{align}} 

The constants $b_i(\omega,k)$ need to be fixed by imposing boundary conditions. The first boundary condition is that there are no outgoing modes at the event horizon $z=z_h$. Expanding the solution near $z=z_h$ and imposing infalling boundary conditions we can obtain $b_2(\omega,k)$ in terms of $b_1(\omega,k)$. Putting it all together we get,

{\scriptsize
\begin{align}
    \phi(z, \omega,k)  & =  b_1(\omega,k) \Gamma \left(1-2 \nu\right) z^{\frac{1}{2}+\nu} (z-z_h)^{\frac{i \omega  z_h}{r_h}} \left( z^{-2\nu} \, _2F_1\left(\frac 1 2  - \nu  - \frac{i q}{\sqrt 2 d},\frac 1 2  - \nu +\frac{i q}{\sqrt 2 d}  + \frac{ i  \omega }{2 \pi T^t};1-2 \nu ;\frac{z}{z_h}\right) \right . \nonumber \\
    & \! \! \left.  -  \frac{z_h^{-2 \nu} \Gamma \left( \frac 1 2 + \nu + \frac{i q }{\sqrt 2 d}\right)   \Gamma \left(\nu -  \frac{i q}{\sqrt{2}d} +\frac{1}{2}-\frac{ i  \omega }{2 \pi T^t}\right)}{  \Gamma \left(\frac 1 2 - \nu + \frac{i q }{\sqrt 2 d}\right) \Gamma \left(\frac 1 2 -\frac{i q}{\sqrt{2}d} - \nu -\frac{ i  \omega }{2 \pi T^t}\right)} \, _2F_1\left(\frac 1 2  + \nu  - \frac{i q}{\sqrt 2 d},\frac 1 2  + \nu +\frac{i q}{\sqrt 2 d}  +\frac{ i  \omega }{2 \pi T^t};1+2 \nu ;\frac{z}{z_h}\right) \right), \label{nearsol1}
\end{align}}
where,
\begin{align}
    \nu = \frac{1}{d}\sqrt{m^2 - \frac{q^2}{2} + \frac{k^2}{r_h^2} + \frac{d^2}{4} }. \label{nus}
\end{align}
The second boundary condition comes from matching the far-away solution (\ref{farphi}) with the near-horizon solution (\ref{nearsol1}) in the middle region. This fixes the constant $b_1(\omega,k)$ in terms of the constants $c_2(\omega,k)$ and $c_1(\omega,k)$. 

Since the near-horizon geometry is itself asymptotically $AdS_2$ (times $\mathbb R_{d-1}$), we can do the following manipulation \cite{Faulkner:2009wj,Faulkner:2010tq}. Imagine a one dimensional conformal field theory that lives at the boundary of this asymptotically $AdS_2$ geometry. We can use holography for this $AdS_2/CFT_1$ system and obtain the two-point function of scalar operators in this $CFT_1$. Since this system arises in the deep IR of the full $AdS_{d+1}/CFT_d$ system, we refer to the $CFT_1$ two-point function as the IR Green's function. As before, the two-point function can be obtained by expanding (\ref{nearsol1}) near the asymptotic $AdS_2$ boundary $z \rightarrow 0$, and evaluating the ratio of the normalizable mode to the non-normalizable mode. Doing that we obtain,
\begin{align}
    G_{IR}(\omega,k) = -  T^{2\nu}  \frac{\Gamma \left(1-2 \nu \right)  \Gamma \left(\frac 1 2+ \nu + \frac{i q }{\sqrt 2 d}\right) \Gamma \left(\frac{1}{2}-\frac{ i  \omega }{2 \pi T} + \nu - \frac{i q}{\sqrt 2 d}\right)}{\Gamma \left(1+2 \nu \right) \Gamma \left(\frac 1 2 -\nu + \frac{i q} {\sqrt 2 d}\right) \Gamma \left(\frac{1}{2}-\frac{ i  \omega }{2 \pi T} -\nu - \frac{i q}{\sqrt 2 d}\right)} \left( \frac{4 \pi }{\kappa r_h} \right)^{2 \nu}. \label{gir}
\end{align}
As shown in the appendix, the IR Green's function for the fermionic case is quite similar but with, 
\begin{align}
    \nu = \frac{1}{d}\sqrt{m^2 - \frac{q^2}{2} + \frac{k^2}{r_h^2}  }. \label{nuf}
\end{align}
% Using $G_{IR}$ we can rewrite our near-horizon solution (\ref{nearsol1}) as,
% \begin{align}
%     \phi(z,\omega,k) = b_1(\omega,k)
% \end{align}
Having obtained the near-horizon solution (\ref{nearsol1}) and far-away solution (\ref{farsol}), we will now match them in the following section.

%%%%%%%%%%%%%%%%%%%%%%%%%%%%%%%%%%%%%%%%%%%%%%%%%%%%%%%%%%%%%%%%%%%%%%%%%%%%%%%%%%%%%%%%%%%%%%%%%%%%%%%%%
%%%%%%%%%%%%%%%%%%%%%%%%%%%%%%%%%%%%%%%%%%%%%%%%%%%%%%%%%%%%%%%%%%%%%%%%%%%%%%%%%%%%%%%%%%%%%%%%%%%%%%%%%
%%%%%%%%%%%%%%%%%%%%%%%%%%%%%%%%%%%%%%%%%%%%%%%%%%%%%%%%%%%%%%%%%%%%%%%%%%%%%%%%%%%%%%%%%%%%%%%%%%%%%%%%%
%%%%%%%%%%%%%%%%%%%%%%%%%%%%%%%%%%%%%%%%%%%%%%%%%%%%%%%%%%%%%%%%%%%%%%%%%%%%%%%%%%%%%%%%%%%%%%%%%%%%%%%%%
%%%%%%%%%%%%%%%%%%%%%%%%%%%%%%%%%%%%%%%%%%%%%%%%%%%%%%%%%%%%%%%%%%%%%%%%%%%%%%%%%%%%%%%%%%%%%%%%%%%%%%%%%%%%%%%%%%%%%%%%%%%%%%%%%%%%%%%%%%%%%%%%%%%%%%%%%%%%%%%%%%%%%%%%%%%%%%%%%%%%%%%%%%%%%%%%%%%%%%%%%%%%%%%%%%

\section{Matched asymptotics for scalars} \label{asympsc}
In order to match the far-away solution with the near-horizon solution, we first need to understand the regime of validity of the individual solutions. The far-away solution relied on the assumption that $f(r) = 1 $ at leading order in $d$. In terms of the mid-region coordinates  defined in (\ref{coord}) this corresponds to,
\begin{align}
    \rho \gtrsim d. \label{farval}
\end{align}
While the near-horizon $AdS_2$ region is valid in a region,
\begin{align}
    \rho -1 \lesssim \frac{1}{d^{n_1}}, \label{nearval}
\end{align}
%\begin{align}
 %   f(r) = 1 + O\left(\frac{1}{d^{n_1}} \right), \label{far1}
%\end{align}
where $n_1$ is some $O(1)$ positive number. Clearly, there is no overlapping range of validity between (\ref{farval}) and (\ref{nearval}). To match our two solutions we will thus need to solve the bulk equation of motion in the mid-region geometry. The equation in the middle region is again a second order differential equation that comes with two integration constants. These get fixed by matching the mid-region solution with the near-horizon solution (\ref{nearsol1}) at small $\rho-1$, and with the far-away solution (\ref{farphi}) at large $\rho$. 

We will find that for certain parameter regimes, we would be able to directly match the far-away solution with the near-horizon solution without solving for the middle region. We will analyse different parameter regimes separately since they give rise to different physics as we now elaborate.
% While the near solution is valid only in the near-horizon limit. Although the near-horizon limit corresponds to $z \sim O(1)$, it can be easily checked that the near-horizon equations of motion form a perturbative expansion in $\frac{\omega}{d z}$. Thus we can extrapolate the near-horizon solution beyond $z_h \geq z \gtrsim 1$ to, 
% \begin{align}
% z_h \geq z \gtrsim \frac{1}{d^{n_2}}, \label{near1}
% \end{align}
% where $n_2$ is strictly smaller than one and greater than zero. In terms of the radial coordinate $r$ this region corresponds to  $f(r) = 1 + O\left(\frac{1}{d^{n_2}} \right)$.  Thus we can match the near-horizon and far-away solutions at leading order consistently. Note that this analysis is true for generic values of $\nu$. In later sections, we will see that while tuning $\nu$ to find bulk Fermi surfaces, we need to adapt a more sophisticated matching scheme.
Recall from the previous section that $\nu$ for scalars is given by,
\begin{align}
     \nu = \frac{1}{d}\sqrt{m^2 - \frac{q^2}{2} + \frac{k^2}{r_h^2} + \frac{d^2}{4} }, \label{nu1}
\end{align}
We see that $\nu$ can take any value depending on the values of the mass, charge and the spatial momenta of the scalar field. We can classify these values into three possible regimes,
\begin{align}
    i) \, \nu \sim 1,  \quad  \quad ii) \, \nu \sim \frac{1}{ d^{n_2}  }, \quad  \quad iii) \, \nu \sim d^{n_3},
\end{align}
where $n_2,n_3$ are some $O(1)$ positive numbers. 

Of the three, we will not consider the case where $\nu$ increases with $d$. The reason is that we are eventually interested in a parameter regime where the two-point function takes the form (\ref{mit}). It is well known (as is also shown below) that such a form is possible only if there exists a range of momenta $k$ for which (\ref{nu1}) could become imaginary. If we assume that $m$ and $q$ have the same large $d$ scaling (or to be more precise, $m$ grows at least as fast as $q$ in the large $d$ limit), then the scenario where $\nu$ increases with $d$ corresponds to an operator with scaling dimension $\Delta \gg d$ in the dual $CFT_d$. This is a highly irrelevant operator, and is not of interest to us.

%This requires us to such values of $\nu$ correspond to $\Delta > d$ in the boundary $CFT_d$. Since our focus is only on relevant or marginal deformations on the boundary, we will work with the first two cases where $\nu$ is $O(1)$ or is $O\left(1 / d^{n_2}\right)$. 

In the rest of the section we will work with the case when $\nu$ is $O(1)$, and $m,q,k$ are also $O(1)$. The scenario where $\nu \sim 1$, but $m,q$ and $k$ take generic $O(d)$ values, is quite similar as shown in appendix \ref{largenu}. %$\nu$ can be $O(1)$ if $m,q$ and $k$ take $O(1)$ values, or if they take generic $O(d)$ values. 
If $m,q$ or $k$ take fine tuned $O(d)$ values, $\nu$ could be made arbitrarily small by cancelling the $d^2/4$ inside the square root (\ref{nu1}). This fine tuned case precisely corresponds to finding a UV two-point function that takes the form (\ref{mit}), as shown in section \ref{find}. %The case where $\nu \sim O(1)$, and $m,q,k$ take generic $O(d)$ values gives the same physics as shown in 

Before we begin with the $\nu \sim 1$ matching problem, we would like to explain the qualitative nature of the solution in our geometry. The far-away solution when expanded near the mid-region takes the following form,
\begin{align}
    \phi(r,\omega,k) = c_+(\omega,k) z^{\frac{1}{2} + \nu } + c_-(\omega,k) z^{\frac{1}{2} - \nu }, \label{m1}
\end{align}
where $c_{\pm}$ are some linear combinations of $c_1$ and $c_2$ mentioned in (\ref{farphi}). Let that be given by,
\begin{align}
    c_2(\omega,k) & = b^+_2(\omega,k) c_+(\omega,k) +b^-_2(\omega,k) c_-(\omega,k), \nonumber \\  c_1(\omega,k) & = b^+_1 (\omega,k)c_+(\omega,k) +b^-_1(\omega,k) c_-(\omega,k),
\end{align}
where the coefficients $b_i^{\pm}(\omega,k)$ are obtained by the explicit matching calculation.
Matching (\ref{m1}) with the near-horizon solution imposes that, 
\begin{align}
    \frac{c_+(\omega,k)}{c_-(\omega,k)} \equiv G_{IR}(\omega,k).
\end{align}
The full UV two-point function (\ref{farsol}) is given by the ratio of $c_i(\omega,k)$. Putting the above together we obtain,
\begin{align}
     \frac{c_2(\omega,k)}{c_1(\omega,k)} & = \frac{b^+_2(\omega,k) c_+(\omega,k) +b^-_2(\omega,k) c_-(\omega,k)}{b^+_1 (\omega,k)c_+(\omega,k) +b^-_1(\omega,k) c_-(\omega,k)}, \nonumber \\
    & = \frac{b^+_2(\omega,k) G_{IR}(\omega,k) +b^-_2(\omega,k) }{b^+_1 (\omega,k) G_{IR}(\omega,k) +b^-_1(\omega,k) }. \label{bulk}
\end{align}
This is the most general form of the UV two-point function, that only depends on the fact that the bulk equation of motion is a second order differential equation.

We begin by first expanding the far-away solution (\ref{farphi}) near the mid-region. The radial coordinate $r$ can be expressed in terms of the middle region coordinate as,
\begin{align}
    r= r_h \left(1 + \frac{\log \rho}{d}\right).
\end{align}
Using the identities (\ref{besexp}) for expansion of Bessel functions at asymptotically large orders, we expand the far-away solution (\ref{farphi}) near the middle region to obtain,

{\scriptsize
\begin{align}
    \phi\left(\rho,\omega,k\right)   = &  \frac{ N  }{ \left(2\pi d \nu r_h^d \rho \right)^{\frac{1}{2}} } \left(  \frac{ c_1(\omega,k) \Gamma \left(1 - \sqrt{\frac{d^2}{4} + m^2}\right) \cos\left( \pi \sqrt{ m^2+\frac{d^2}{4}}  \right)+c_2(\omega,k)\Gamma \left(1 + \sqrt{\frac{d^2}{4} + m^2}\right) e^{i \pi \sqrt{ 4m^2+d^2}}}{\rho^{ \nu}  e^{\alpha}} \right. \nonumber \\
    & \left . - 2 \rho^{ \nu}   e^{\alpha}  c_1(\omega,k) \Gamma \left(1 - \sqrt{\frac{d^2}{4} + m^2}\right) \sin\left( \pi \sqrt{ m^2+\frac{d^2}{4}} \right)  \right), \label{farnear}
\end{align}}
where,
\begin{align}
    N & \equiv 2^{-\frac{3 d}{4}} i^{- \sqrt{4 m^2+d^2}}  \left(q^2 r_h^2 - 2 k^2 + \frac{2 \sqrt{2} q r_h \omega}{d} + \frac{2 \omega ^2}{d^2}\right)^{d/4}, %\quad \theta \equiv \pi \sqrt{4 m^2+d^2}, 
    \nonumber \\
    \alpha & \equiv  \sqrt{\frac{q^2}{2}-\frac{ k^2}{r_h^2}+\nu^2 d^2 }  \log \left(\frac{\nu d  +  \sqrt{\frac{q^2}{2}- \frac{k^2}{r_h^2}+\nu^2 d^2 }}{\sqrt{\frac{q^2}{2}-\frac{k^2}{r_h^2}}}\right) -  \nu d \left(\frac{\sqrt{2} q r_h \omega }{q^2 r_h^2 -2 k^2}+1\right) .
\end{align}

We will match this with the near-horizon solution (\ref{nearsol1}) expanded near the middle region. Using (\ref{coord}), (\ref{tempco}) and (\ref{zh}) the $\rho$ coordinate can be written in terms of $z$ as,
\begin{align}
    \rho = 1 + \frac{z_h}{z} \left( \frac{\uu}{d} -1 \right).
\end{align}
The limit $z \rightarrow 0$ in the near-horizon geometry corresponds to going towards the mid-region. We take this limit such that $\rho \sim \frac{z_h \uu}{z d} \rightarrow \infty$ i.e we take $z\rightarrow 1/d^{n_4}$ where $n_4$ is some number $n_4>1$. This lies far outside the regime of applicability of the near-horizon geometry (\ref{nearval}). However it can be shown that this limit can be taken consistently to match the far-away with the near-horizon solution. We have checked this by explicitly solving the bulk equation of motion perturbatively in $1/d$ in the middle region. In the following section, we will find that for small $\nu$, it is necessary to solve the bulk equation of motion in the middle region before we can match the far-away solution with the near-horizon solution. Taking the limit $z\rightarrow 0$, and using the asymptotic expansions of the hypergeometric function and the IR Green's function (\ref{gir}), the near-horizon solution (\ref{nearsol1}) becomes,
\begin{align}
    \phi(z,\omega,k) = b_1(\omega,k) (-z_h)^\frac{i z_h \omega}{r_h} \left( z^{\frac{1}{2}-\nu} -  \frac{z^{\frac{1}{2} +\nu }}{z_h^{2\nu}} G_{IR}(\omega,k)\right).  \label{nearfar}
\end{align}
%Matching this with (\ref{farnear}) will give us the ratio $c_2/c_1$ 

We can now match our two solutions (\ref{farnear}) and (\ref{nearfar}). Up to exponentially small corrections in $d$ we get,
\begin{align}
    \frac{c_2(\omega,k)}{c_1(\omega,k)} =   G_{IR}(\omega,k)  \left(\frac{\kappa e^{- i \pi  d } }{d}\right)^{2 \nu }.
\end{align}
Thus the full two-point function when $\nu \sim 1$ is given by,
\begin{align}
    G_{UV} (\omega,k) =  2^{\frac{3}{2} (d-2 \Delta )} \left(-2 k^2+q^2 r_h^2+2 \sqrt{2} q r_h \omega +2 \omega ^2\right)^{\Delta -\frac{d}{2}}   G_{IR}(\omega,k)  \left(\frac{\kappa e^{- i \pi  d  } }{d}\right)^{2 \nu }. \label{nu1match}
\end{align}
This is quite different than the answer advertised in (\ref{mit}). In particular, the UV Green's function does not have a non-trivial pole for $\omega \rightarrow 0, k \rightarrow k_F$. As a result such a parametric range of $\nu$ is of no interest to us. In the next section, we will find a Green's function precisely of the form (\ref{mit}) when $\nu$ is small in the large $d$ limit. %We will also provide an intuitive explanation why the Green's function found here for $\nu \sim 1$ does not take such a form.

Let us now understand why (\ref{nu1match}) does not take the general form (\ref{bulk}) for the two-point function. If we compare (\ref{bulk}) with (\ref{nu1match}) we conclude that $b_1^+$ and $b_2^-$ are zero, or at the least exponentially small in $d$ for the $\nu \sim 1$ matching problem. Qualitatively this is because $b_1^+(\omega,k)$ corresponds to the contribution of the $AdS_2$ normalizable mode to the $AdS_d$ non-normalizable mode, while $b_2^-(\omega,k)$ is the contribution of the $AdS_2$ non-normalizable mode to the $AdS_d$ normalizable mode. These contributions are exponentially small in $d$ since a mode that is normalizable or non-normalizable in the mid-region geometry stays so throughout the entire geometry, as can be seen from the form of the scalar potential in figure \ref{nu1fig} of appendix 
\ref{largenu}. It can be explicitly seen from (\ref{comp}), that the exponentially small contributions become polynomial precisely when $\nu \sim 1/d^{1/3}$. 

Intuitively when $\nu$ takes such a form, there is no distinction between the normalizable mode $z^\nu$ and the non-normalizable mode $z^{-\nu}$ in the near-horizon geometry, and the order of magnitudes of the matching coefficients $b_i^{\pm}(\omega,k)$ in terms of large $d$ counting are the same. In that case, we do obtain a Green's function that takes the form given in (\ref{bulk}), as we show in the following section.

%We will give an intuitive explanation why this is the case, and show that for small $\nu$ we do recover (\ref{mit}) in the following section. 

%What we find is that the ratio is simply given by the 
%Since $m,q,k$ are $O(1)$, $\nu$ at leading order is just given by $\frac{1}{2}$

% Before we expand the far-away solution in the near-horizon region, we would like to understand what is considered to be a generic/natural value of $\nu$. This requires us to answer what are the natural values of $m,q$ and $\vec k$. Consider the boundary operator dimension of a scalar operator given by (\ref{scdim}). We find that unless  In a field theory with $d$ space time dimensions, it is natural to assume that $k\equiv |\vec{k}|$ scales linearly with 

%%%%%%%%%%%%%%%%%%%%%%%%%%%%%%%%%%%%%%%%%%%%%%%%%%%%%%%%%%%%%%%%%%%%%%%%%%%%%%%%%%%%%%%%%%%%%%%%%%%%
%%%%%%%%%%%%%%%%%%%%%%%%%%%%%%%%%%%%%%%%%%%%%%%%%%%%%%%%%%%%%%%%%%%%%%%%%%%%%%%%%%%%%%%%%%%%%%%%%%%%
%%%%%%%%%%%%%%%%%%%%%%%%%%%%%%%%%%%%%%%%%%%%%%%%%%%%%%%%%%%%%%%%%%%%%%%%%%%%%%%%%%%%%%%%%%%%%%%%%%%%
%%%%%%%%%%%%%%%%%%%%%%%%%%%%%%%%%%%%%%%%%%%%%%%%%%%%%%%%%%%%%%%%%%%%%%%%%%%%%%%%%%%%%%%%%%%%%%%%%%%%
%%%%%%%%%%%%%%%%%%%%%%%%%%%%%%%%%%%%%%%%%%%%%%%%%%%%%%%%%%%%%%%%%%%%%%%%%%%%%%%%%%%%%%%%%%%%%%%%%%%%

\subsection{Finding bulk Fermi surfaces} \label{find}
In this section, we will focus on the case when $\nu$ is fine tuned to be a small number that vanishes in the large $d$ limit. %As shown in the previous section, the full UV two-point function is rather uninteresting when $\nu$ is $O(1)$. 
In particular, we will show that if,
\begin{align}
    \nu = \frac{V(m,q,k_F)}{d^{n_2}},   \label{nusc}
\end{align}
where $V(m,q,k_F)$ and $n_2$ are some positive $O(1)$ numbers, the UV Green's function has a pole as $\omega \rightarrow 0, k \rightarrow k_F$. 
The value of $n_2$ is different for scalars and fermions, and depends on what we assume to be the natural large $d$ scaling for the momenta $k\equiv \sqrt{k^ik_i}$. In a $d+1$ dimensional spacetime theory, if we assume that the individual momentum components are $O(1)$ numbers in some units, it follows that the momentum $k$ is an $O(\sqrt d)$ number in those units. This is quite similar to the rescaling in the fermion hopping matrix $t_{ij} \rightarrow \frac{t}{\sqrt d}$ that is done in DMFT to ensure that a well-defined $d \rightarrow \infty$ limit exists. We will thus work with the case where $k \sim \sqrt d$, and find that for $m, q \sim \sqrt d$, the UV two-point function for fermions has a non-trivial zero for $n_2 = 2/3$. 

Similar arguments hold true for scalars and we find that the UV two-point function has a non-trivial zero for $n_2 = 2/3$. However, in our holographic setup there is an additional parameter regime that is natural for scalars but not for fermions. This difference stems from the fact that a massless scalar field corresponds to a marginal operator of scaling dimension $\Delta = d$ in the boundary $CFT_d$, while a massless fermionic field is dual to an operator with scaling dimension $\Delta = \frac{d}{2}$. In terms of the scaling exponent $\nu$, this corresponds to the difference of $d^2/4$ between the scalars (\ref{nus}) and the fermions (\ref{nuf}). Due to this form of the scaling exponent for scalars (\ref{nus}), there are two interesting parametric regimes. The first, where $k \sim \sqrt d$ like the fermions discussed above. The second, where $k \sim d$. We will not work out the case when $k \sim \sqrt d$ for scalars, since the results are identical to the fermionic case discussed in appendix \ref{fermat}. In the rest of the section we will deal with the case where $m,q,k \sim d$ for scalars, and find that the UV two-point function has a non-trivial zero for $n_2=1/3$. 

%We will deal with fermions , and show that $n_2=2/3$.

%As can be seen from (\ref{nu1}), a small value of $\nu$ can only be obtained if $m,q$ and $k$ take fine tuned values of order $d$. This is the naturalThe fermions have a different parametric regime of interest, ,In principle $m,q$ or $k$ could take values that scale differently with $d$. For the case of fermions we will see that $\nu$ can be made small without taking $m,q$ and $k$ to be of order $d$, as shown in appendix \ref{fermat}. This difference arises from the fact in $AdS/CFT$ a massless scalar in the bulk corresponds to a marginal operator on the boundary, while a massless fermion corresponds to a relevant operator as elaborated in appendix \ref{fermat}.

The faraway solution (\ref{farphi}) and the near-horizon solution (\ref{nearsol1}) still remain the same, but the details of the matching calculation change a bit for the following reasons. The first is that when $\nu$ is small, the order and the argument of the Bessel functions in (\ref{farphi}) are both large, and approximately equal to each other. This requires a new asymptotic expansion of the Bessel function, and we get a new expression for the far-away solution when expanded near the mid-region as shown near (\ref{farmid}) in the appendix. The second is that we also need to solve the bulk equations of motion in the mid-region to match the far-away solution with the near solution. We perform this matching explicitly in appendix \ref{finetu}, and only focus on the qualitative features here. 

As argued near (\ref{bulk}) the most general solution to the matching problem is given by,
\begin{align}
      \frac{c_2(\omega,k)}{c_1(\omega,k)} & = \frac{b^+_2(\omega,k) G_{IR}(\omega,k) +b^-_2(\omega,k) }{b^+_1 (\omega,k) G_{IR}(\omega,k) +b^-_1(\omega,k) }.
\end{align}
If $b^-_1(\omega,k)$ has a non-trivial zero such that,
\begin{align}
    b_1^-(\omega,k) \equiv \omega + v_F (k-k_F) + \ldots ,
\end{align}
where the $\ldots$ refer to higher order terms in $\omega$ or $k-k_F$, the UV two-point function can be written as,
\begin{align}
    \frac{c_2(\omega,k)}{c_1(\omega,k)} & = \frac{b_2^-(0,k_F)}{\omega + v_F(k-k_F) + b_1^+ (0,k_F) G_{IR} (\omega,k_F)}. \label{bulkg}
\end{align}
This is the main result of this section. We find that the full UV two-point function takes the form of the quasiparticle Green's function near a Fermi surface with the IR Green's function serving the role of the self energy. While this fact has been known since the original work of Faulkner et al. \cite{Faulkner:2009wj}, the point of our work is to emphasize that at large $d$ the matching calculation can be done explicitly order by order in a large $d$ expansion. Using which we found the large $d$ scaling of $\nu$ (\ref{nusc}), and a closed form expression for evaluating $V(m,q,k_F)$ as given by (\ref{zero}) for the scalars and (\ref{ferai}) for the fermions. 

As mentioned in footnote \ref{bose}, the two-point function for bosons signals an instability. This is because the two-point function can be written as (\ref{bulkg}), only when there exists a range of momenta where $\nu$ (\ref{nu1}) can become imaginary. This requires that the charge of the scalar field is greater than its mass $q \geq \sqrt{2 m^2 + \frac{d^2}{2}} $. It is well known  that such a charged scalar field in a charged black brane geometry leads to an instability \cite{Hartnoll:2008kx}. As show in \cite{Faulkner:2009wj}, it is related to super radiance of the black brane in the bulk, and the spin statistics of the dual operator in the boundary. The charged fermion considered in appendix \ref{fermat} does not have this instability.

 The Fermi velocity and the momentum in (\ref{bulkg}), as shown in appendix \ref{finetu}, are given at leading order in $d$ by,
\begin{align}
    v_F & = \frac{\sqrt{ q^2-2 m^2- \frac{d^2}{2}}}{ q }, \qquad k_F  = r_h\sqrt{ \frac{q^2}{2}- m^2- \frac{d^2}{2}}.
\end{align}
Note that $m$ and $q$ both scale linearly with $d$, thus $v_F$ is an $O(1)$ number while $\frac{k_F}{\mu}$ is $O(d)$, where $\mu$ is given by (\ref{tmu2}). The self energy is given by,

{\scriptsize
\begin{align}
    \Sigma(\omega,k) & \equiv  b_1^+(0,k_F)  G_{IR} (\omega,k_F), \nonumber \\
    & = b_1^+(0,k_F)  T^{\frac{2V(m,q,k_F)}{d^{1/3}}}  \left( 1-\frac{4 V(m,q,k_F)\left( \psi ^{(0)}\left(-\frac{i q}{\sqrt{2} d}+\frac{1}{2}-\frac{i  \omega }{2 \pi  T}\right)+ \psi ^{(0)}\left(\frac{i q}{\sqrt{2} d}+1\right)+4 \gamma + \log\left( \frac{ r_h}{4} \right) \right)}{d^{1/3}} \right), \label{self}
\end{align}}
where $\psi^{(0)}$ is the polygamma function and $\gamma$ is the Euler constant. We have used the $AdS_2$ Green's function (\ref{gir}), and have expanded to the first subleading order in $d$ to obtain (\ref{self}). The constants $ b_1^+(0,k_F)$ and $b_2^-(0,k_F)$ are $O(1)$ numbers, with at the most a logarithmic dependence on $d$, and are given by (\ref{beesol}) in the appendix. As long as the temperature and frequency are not exponentially small in $d$, we can also expand the exponent in the self-energy (\ref{self}). Up to shifts and redefinition of the Fermi momenta, we obtain that,
\begin{align}
    \text{Re} \, \Sigma \, \propto \, \frac{V(m,q,k_F)}{d^{1/3}} \log T , \qquad  \text{Im} \, \Sigma \, \propto \, \frac{V(m,q,k_F)}{d^{1/3}} . \label{reim}
\end{align}
The scaling exponent $V(m,q,k_F)$ can be obtained by finding the zero of a certain combination of Airy functions as argued near (\ref{zero}), and has at the most a logarithmic dependence on $d$. The presence of Airy functions has a very intuitive explanation that we now elaborate.

%The self energy in (\ref{bulkg}) is given by the first concrete prediction that we find is that in this large $d$ setup, the coefficient $b_1^-(\omega,k)$ has a non-trivial zero only when, 
%\begin{align}
%\nu(m,q,k) = \frac{V(m,q,k)}{d^{\frac 1 3}}, \label{mainnu}
%\end{align}
%where $V(m,q,k)$ is an $O(1)$ number. This means that the UV two-point function (\ref{bulk}) takes the form (\ref{bulkg}) only when the scaling exponent $\nu$ has a large $d$ scaling that is of order (\ref{mainnu}).. %The coefficients $b_i(\omega,k_F)$ are complicated functions in the variables $m,q,k_F$ that are given in the appendix. 
%The coefficients  and $b_1^+(0,k_F)$ 

% Recall that the IR Green's function is given by,
% \begin{align}
%     G_{IR}
% \end{align}

\begin{figure}
    \centering
    \includegraphics[scale=0.45]{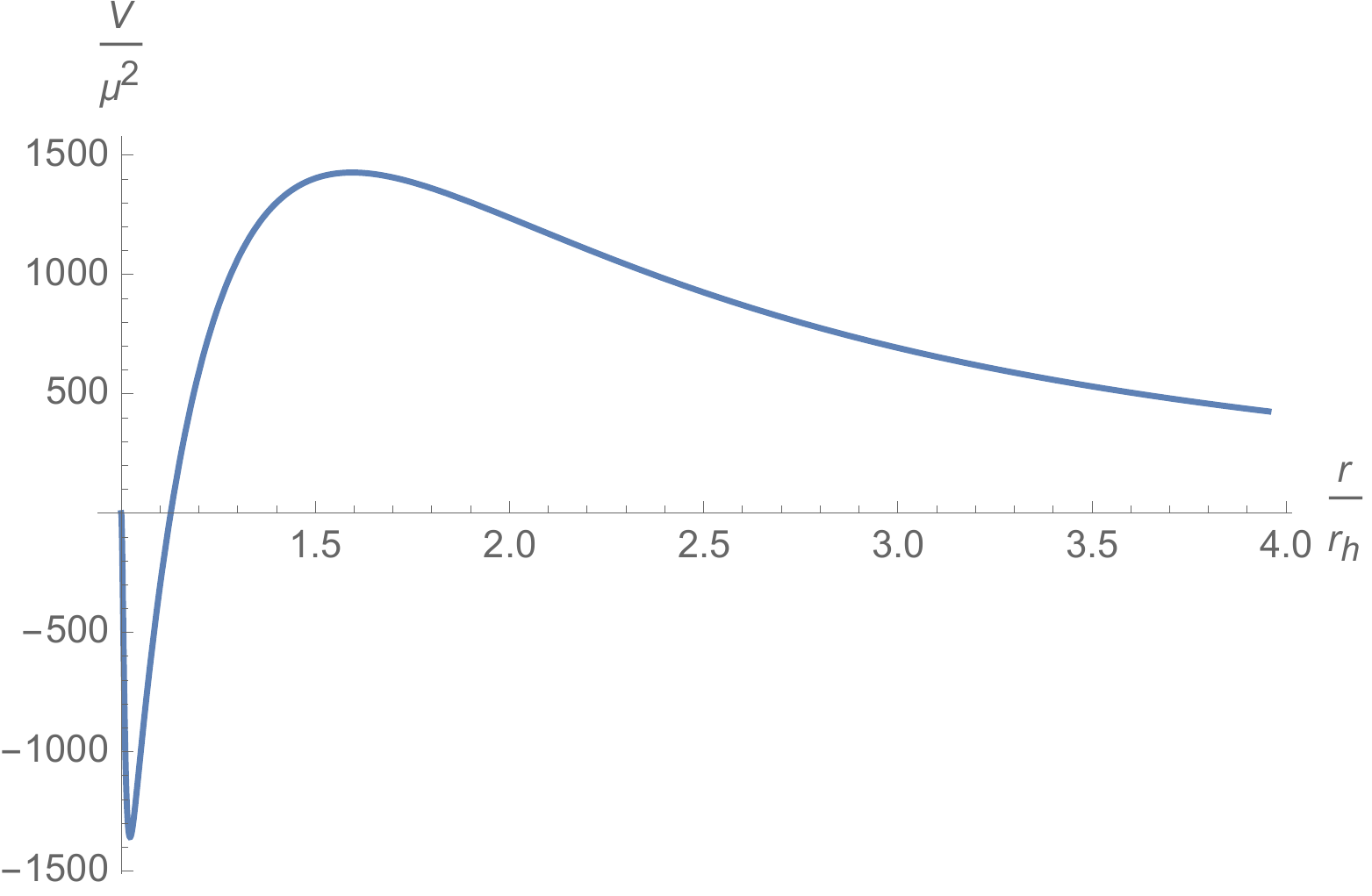}
    \includegraphics[scale=0.45]{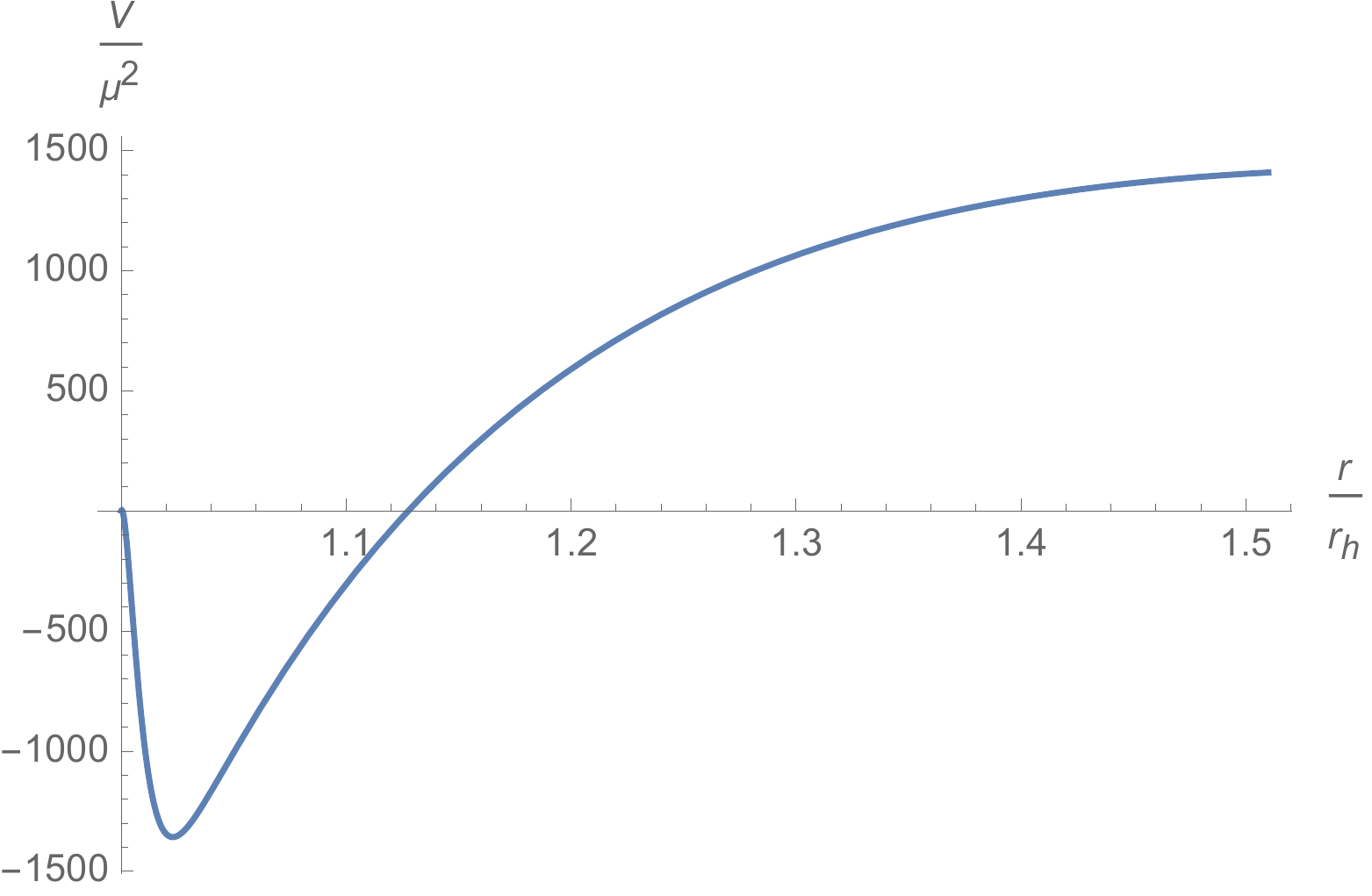}
    \caption{The left panel is a plot of the potential for fine tuned $\nu$. The right panel is a closeup of the potential near the horizon, which looks like a triangular well.}
    \label{tunsc}
\end{figure}

 Consider the mode that grows like $z^{\frac{1}{2} -\nu }$ in the near-horizon region as given in (\ref{m1}). In the process of finding a bulk UV two-point function that takes the form (\ref{bulkg}), we have imposed two boundary conditions on this mode. The first is the boundary condition in the near-horizon region where we match it with (\ref{nearfar}). The other is requiring that $b^-_1$ is zero as $\omega \rightarrow 0$ for some $k=k_F$. In other words, we are looking for a normalizable solution of a second order differential equation when $\omega=0$. This is just the Schrodinger bound state problem in quantum mechanics. As discussed near (\ref{sceq}), the bulk Klein-Gordon equation can be written as a Schrodinger equation using a change of variables and a field redefinition,
\begin{align}
    \frac{d^2 \Phi(\sigma)}{d\sigma^2} - V(\sigma) \Phi(\sigma) =0.
\end{align}
For generic $\nu$ the potential was shown in figure \ref{scpot}. For fine tuned values of $\nu$ i.e. when $\nu = \frac{V(m,q,k)}{d^{1/3}}$ the potential at large $d$ becomes what is shown in figure \ref{tunsc}.
The potential develops an almost triangular well in the near-horizon region at large $d$. Finding bound states in a triangular well is a standard problem, the spectrum of which is obtained by finding the zeros of certain combination of Airy functions, just like in (\ref{zero}). Note that figure \ref{tunsc} is drawn in the $r$ coordinate, but a similar triangular well picture holds true in the $\sigma$ coordinate as well.

\section{Discussion}

In this note, we argued that holographic non-Fermi liquids are amenable to a large $d$ perturbative analysis. We found explicit results for the Fermi velocity, the Fermi momentum, and the scaling exponent $\nu$. We showed that at large $d$, the natural values of $\nu$ are $V(m,q,k_F)/d^{1/3}$ for scalars and $V(m,q,k_F)/d^{2/3}$ for fermions, where $V(m,q,k_F)$ is given by the zeros of a certain combination of Airy functions. For such small values of $\nu$, we found that the self energy (\ref{reim}) has a constant imaginary part proportional to $\nu$, and a real part that is proportional to $\nu \log T$.

Our results have intriguing similarities to other calculations in the recent literature. For instance, a constant quasiparticle lifetime was also found in recent simulations of certain quantum critical metals \cite{Lederer4905}. And a vanishingly small value of $\nu$ also arose in the context of de Haas–van Alphen oscillations in some materials explored recently in \cite{Tan287,Hartnoll:2009kk}. 

At large $d$, we also demonstrated that the self energy of the full two-point function is a function of only $\omega$, for any finite value of $T/\mu$. This is beyond the usual regime of validity found in previous work on holographic non-Fermi liquids, and in fact agrees with the DMFT ansatz. Unlike the DMFT method where the $d \rightarrow \infty$ limit is taken first, our analysis works at all orders in a perturbative series in $1/d$. A possible generalisation of our work is to consider theories with hyperscaling violation \cite{hyper}. Qualitatively, the effective number of dimensions in such theories is $d-\theta$, and a similar analysis can be done at finite $d$ but large and negative hyperscaling violating exponent $\theta$.

The obvious drawback of our work is that we work with large-$N$ holographic CFTs, which are very different from real-life systems. In addition to that, the Fermi surface and the corresponding charge density found in our work are $O(N^0)$ and are subleading compared to the $O(N^2)$ contribution from the black hole. In other words, to explore further properties of this system we would need to do a bulk one-loop calculation. It is unclear whether the large-$d$ expansion behaves well at subleading orders in $N$, and we leave that to future work. 

Finally, it is interesting to note that previous works on the large $N$ 
limits of certain condensed matter systems have similarities with the large $N$ holographic analysis. For instance, the large $N$ multichannel Kondo effect at large number of channels $K$ and Abrikosov fermions $Q$ \cite{kondo} has emergent conformal invariance and an IR Green's function that is quite similar to the $AdS_2/CFT_1$ Green's function given in (\ref{gir}).

%Despite the obvious differences between holographic CFTs and real life systems, we believe that our work provides a new perspective on the effectiveness of DMFT like techniques, and we hope to report on this direction in the future.

%and new results about the scaling exponent $\nu$.  We find these connections quite interesting and we hope to elaborate 

%  oscillations   particular, our results about the smallness of $\nu$ in the large $d$ limit, might provide a ne

\acknowledgments
It is a pleasure to thank Sean Hartnoll, Diego Hofman, Steve Kivelson, Gabi Kotliar, Jorrit Kruthoff, John McGreevy, Michael Mulligan, Sri Raghu and Edgar Shaghoulian for several interesting discussions.  This work was supported by 
the Simons Collaboration on Ultra-Quantum Matter, which is a grant from the Simons Foundation (651440, SK).  It was also supported by 
the Department of Energy under grant DE-SC0020007 and a Simons Investigator Award.

\appendix
\renewcommand{\thesection}{\Alph{section}}

\section{Matching for fine tuned scalars} \label{finetu}
In this section, we will provide the details for the matching calculation for bulk scalars when $\nu$ is fine tuned to be small. We will perform the matching in the middle region that lies in between the near-horizon $AdS_2$ region and the far-away $AdS_{d+1}$ region. Since the middle region is parametrised by the coordinates $(\rho,t)$, we will first rewrite our near-horizon and far-away results in those coordinates. 

The near-horizon solution (\ref{nearsol1}) at the $AdS_2$ boundary i.e $z \rightarrow 0$ is given by,
\begin{align}
    \phi(z,\omega,k) = b_1(\omega,k)  z^{\frac{1}{2} - \nu }+ b_1(\omega,k) G_{IR}(\omega,k) z^{\frac{1}{2} + \nu}. 
\end{align}
Using (\ref{coord}),
\begin{align}
     \rho  =  \frac{\uu}{d} \left(\frac{z_h }{z } -1 \right) + 1 .
\end{align}
we can express the near-horizon solution in terms of the coordinates $(\rho,t)$,

{\small
\begin{align}
      \phi(\rho,\omega,k) = &  b_1(\omega,k)  \left( \frac{G_{IR}(\omega,k)+1}{d^{\frac 1 2 }}\right) \sqrt{\frac{\kappa}{\rho -1}} + b_1(\omega,k) \left( \frac{G_{IR}(\omega,k)-1}{d^{\frac 5 6}} \right) V \sqrt{\frac{\kappa}{\rho -1}} \log \left(\frac{\kappa}{d (\rho -1)}\right)  \nonumber \\
      &  + \frac{1}{2} b_1(\omega,k)   \left( \frac{G_{IR}(\omega,k)-1}{d^{\frac 7 6}} \right) V^2 \sqrt{\frac{\kappa}{\rho -1}} \log ^2\left(\frac{\kappa}{d (\rho -1)}\right) + \ldots ,\label{nearmid}
\end{align}}
where we have used $\uu z_h = d\lambda = \kappa$.

Similarly, the far-away solution (\ref{farphi}) can be expanded in the mid-region by using the relation, 
\begin{align}
    r= r_h \rho^{\frac{1}{d-2}} = r_h\left( 1 + \frac{\log \rho}{d} + \ldots \right).
\end{align}
The Bessel functions in (\ref{farphi}) have a rich set of asymptotic expansions, depending on the relative magnitudes of their order and argument. When $\nu$ is of order $1/d^{\frac 1 3}$, it can be easily checked that the argument and the order both grow linearly in $d$, while their difference is of order $d^{\frac 1 3}$. When this is the case, we can use the beautiful result of Olver \cite{olver_1952} given in equation (\ref{hailolver}) to write,

{\scriptsize
\begin{align}
    \phi(r,\omega,k) = &  \frac{ \mathcal N_1  2^\frac{1}{3} }{r_h^{d/2}} \left[ \frac{2^{1/3}  \left(( \tilde c_1 \cos \theta i^{2 (d-\Delta )} + \tilde c_2 e^{i \pi  \Delta }) \text{Ai}\left(\frac{2^{2/3} V^2}{\left(4 M^2+1\right)^{2/3}}\right)- \tilde c_1 i^{2 (d-\Delta )} \sin \theta \text{Bi}\left(\frac{2^{2/3} V^2}{\left(4 M^2+1\right)^{2/3}}\right)\right)}{ d^\frac{1}{3} (4 M^2+1)^\frac{1}{6} \sqrt{\rho}} \right. \nonumber \\
     & + \left .   \frac{   \left( \frac{2k_{\perp}}{r_h} \sqrt{2 Q^2-4 M^2-1}+\left(4 M^2 +1\right) \log \rho -2 \sqrt{2} Q \frac{\omega }{r_h} \right) }{d^{2/3} \left(4 M^2+1\right)^{5/6} \sqrt{\rho} } \right. \nonumber \\ 
     & \times \left. \left( (\tilde c_1 \cos \theta i^{2 (d-\Delta )} + \tilde c_2 e^{i \pi  \Delta }) \text{Ai}'\left(\frac{2^{2/3} V^2}{\left(4 M^2+1\right)^{2/3}}\right) - \tilde c_1 i^{2 (d-\Delta )} \text{Bi}'\left(\frac{2^{2/3} V^2}{\left(4 M^2+1\right)^{2/3}}\right) \sin \theta \right) + O\left( \frac 1 d \right)\right], \label{farmid}
\end{align}}
where we have defined,
\begin{align}
    \tilde c_1 \equiv c_1 \Gamma\left( 1 + \frac{d}{2} \sqrt{1 + 4M^2} \right), \qquad  \tilde c_2 \equiv c_2 \Gamma\left( 1 - \frac{d}{2} \sqrt{1 + 4M^2} \right),
\end{align}
and $\theta \equiv \frac{d \pi}{2} \sqrt{4 M^2+1}$. In the appendices for scalars we have employed the notation $m=M d, q= Q d$. Having obtained the asymptotic expansions in the mid region, we will now proceed to match them. There are perhaps quicker ways of doing, but we will proceed in a way that we find the cleanest. 

The expansion (\ref{nearmid}) for the near-horizon solution is valid as long,
\begin{align}
    \rho -1 \ll 1. \label{nearcov}
\end{align}
While the expansion (\ref{farmid}) for the far-away solution is valid when,
\begin{align}
    \rho \gtrsim d. \label{farcov}
\end{align}
Since there is no overlap between the two regions, we will solve the bulk equation of motion in the middle region and match it with the near-horizon and far-away solutions, perturbatively in $1/d$. Let the middle solution be given by,
\begin{align}
    \phi(\rho,k,\omega) \equiv \chi(\rho,\omega,k) + \frac{1}{d^{\frac{1}{3}}}\chi_1(\rho,\omega,k) + \frac{1}{d^\frac{2}{3}}\chi_2 (\rho,\omega,k) + \ldots .  \label{bulkper}
\end{align}
The powers of $d$ are chosen with hindsight obtained from the rest of the calculation. Given this expansion the bulk equation of motion in the middle region can be solved perturbatively. Each $\chi_i(\rho,\omega,k)$ satisfies a second order differential equation, and hence comes with two integration constants. Since each $\chi_i(\rho,\omega,k)$ also satisfies two boundary conditions, one at the near-horizon/mid-region boundary and the other at the mid-region/far-away boundary, the $\chi_i(\rho,\omega,k)$ are fixed uniquely order by order. We will write down the first few $\chi_i(\rho,\omega,k)$,
\begin{align}
    \chi(\rho,\omega,k) &  = \frac{c_1(\omega,k)}{\sqrt{\rho-1}}+\frac{c_2(\omega,k) \log (\rho-1)}{2 \sqrt{\rho-1}}, \qquad \chi_1 (\rho,\omega,k)  = \frac{c_3(\omega,k)}{\sqrt{\rho-1}}+\frac{c_4(\omega,k) \log (\rho-1)}{2 \sqrt{\rho-1}}, \nonumber \\ 
    \chi_2 (\rho,\omega,k) & = \frac{c_5(\omega,k) V^2 \log ^2(\rho-1)+2 c_5(\omega,k)+ c_6(\omega,k) \log (\rho-1)}{2 \sqrt{\rho-1}}. \label{midexp}
\end{align}
We will match (\ref{midexp}) with the near-horizon expansion (\ref{nearmid}) when $\rho \equiv 1 + \frac{\rho_{IR}}{d^{1/3}}$, and with the far-away expansion (\ref{farmid}) when $\rho \equiv \rho_{UV} d$. These clearly lie in the respective regimes of convergence (\ref{nearcov}) and (\ref{farcov}). The parameters $\rho_{IR}$ and $\rho_{UV}$ are some $O(1)$ numbers. Perhaps the higher order $\chi_i(\rho,\omega,k)$ could also be obtained by a Callan-Symanzik type argument by demanding that the physical quantities do not depend upon the parameters $\rho_{IR}$ and $\rho_{UV}$. In the following, we will set $\rho_{IR} = \rho_{UV}=1$. Note that solving for the higher order terms in (\ref{bulkper}) requires the subleading corrections to the mid-region metric (\ref{drn}). We have not listed those corrections to the metric here, but have incorporated them in our computations wherever they were necessary.

Performing this matching calculation at leading order we obtain,
\begin{align}
    \frac{c_2}{c_1} = \frac{\Gamma\left( 1 - \frac{d}{2} \sqrt{1 + 4M^2} \right)}{\Gamma\left( 1 + \frac{d}{2} \sqrt{1 + 4M^2} \right)} \frac{\mathcal N }{\mathcal D}
\end{align}
where,
{\tiny
\begin{align}
    \mathcal N & =2^{2/3} V^2 \left((G+1) K_{\frac{2}{3}}\left(\frac{4 V^3}{12 M^2+3}\right)-\frac{3 (G-1) \left(4 M^2+1\right) r_h K_{\frac{1}{3}}\left(\frac{4 V^3}{12 M^2+3}\right) \log \left(\frac{\kappa }{d^{2/3}}\right)}{\log (d) \left(4 M^2 r_h+r_h\right)+2 k_\perp \sqrt{-4 M^2+2 Q^2-1}-2 \sqrt{2} Q \omega }\right), \nonumber \\ 
    \mathcal D & = \sqrt{3} \pi  \left(4 M^2+1\right)^{2/3} i^{2 (d-2 \Delta )} \sin \theta  \left(-\frac{3 \sqrt[3]{2} (G-1) \left(4 M^2+1\right)^{2/3} r_h V \log \left(\frac{\kappa }{d^{2/3}}\right) \left(\cot \theta  \text{Ai}\left(\frac{V^2}{\left(2 M^2+\frac{1}{2}\right)^{2/3}}\right)+\text{Bi}\left(\frac{V^2}{\left(2 M^2+\frac{1}{2}\right)^{2/3}}\right)\right)}{2 \sqrt{2} Q \omega-\log d \left(4 M^2 r_h+r_h\right)-2 k_\perp \sqrt{2 Q^2-4 M^2-1} } \nonumber \right. \\ 
    & \hspace{4.5cm} \left. +(G+1) \cot (\theta ) \text{Ai}'\left(\frac{V^2}{\left(2 M^2+\frac{1}{2}\right)^{2/3}}\right) +(G+1) \text{Bi}'\left(\frac{V^2}{\left(2 M^2+\frac{1}{2}\right)^{2/3}}\right)\right). \label{matsol}
\end{align}
}
We can tune $V$ such that the ratio $\mathcal N / \mathcal D$  takes the form (\ref{bulkg}). We find that if $V(m,q,k_F)$ satisfies the following relation,
\begin{align}
    \left(2 M^2+\frac{1}{2} \right)^\frac{1}{3} \log d \, \text{Ai}'\left(\frac{V^2}{\left(2 M^2+\frac{1}{2}\right)^{2/3}}\right)=3 V \text{Ai}\left(\frac{V^2}{\left(2 M^2+\frac{1}{2}\right)^{2/3}}\right) \log \left(\frac{\kappa}{d^{2/3}}\right), \label{zero}
\end{align}
the full two-point function has a non-trivial zero as $\omega \rightarrow 0$ and $k \rightarrow k_F$. The constants that appear in (\ref{bulkg}) can now be easily extracted from (\ref{matsol}) and (\ref{zero}). We find,

{\tiny
\begin{align}
    b_2^- & = -\frac{\left(4 M^2+1\right) r_h \log d}{\sqrt{2} Q}, \nonumber \\ 
    b_1^+ & = \frac{ \left(4 M^2+1\right) r_h i^{2 (d-2\Delta )} \log d \, \tan \theta  \left(\sqrt[3]{4 M^2+1} \log (d) \text{Bi}'\left(\frac{V^2}{\left(2 M^2+\frac{1}{2}\right)^{2/3}}\right)-3  V \text{Bi}\left(\frac{V^2}{\left(2 M^2+\frac{1}{2}\right)^{2/3}}\right) \log \left(\frac{ \kappa }{d^{2/3}}\right)\right)}{ 2^{5/6} Q V \text{Ai}\left(\frac{V^2}{\left(2 M^2+\frac{1}{2}\right)^{2/3}}\right) 6 \log \left(\frac{ \kappa }{d^{2/3}}\right)}. \label{beesol}
\end{align}}
As mentioned in the main text, both constants have at most a logarithmic dependence on $d$. We can integrate (\ref{zero}) to get a simpler equation,
\begin{align}
    \text{Ai} \left( \frac{V^2}{(2M^2 + \frac{1}{2})^{2/3}}\right) = \exp\left( \frac{2 \log \left( \frac{\kappa}{d^{2/3}} \right)}{\log d} \frac{V^3}{2M^2 + \frac{1}{2}}  \right)
\end{align}
The equation has a unique solution, and corresponds to finding a bulk Fermi surface.

\section{Matching for scalars for $\nu \sim 1$ and $m,q,k \sim d$ } \label{largenu}
In this section we will perform the matching calculation for scalars in a parameter regime that wasn't discussed in the main text. We argued in section \ref{find} that the UV two-point function has a non-trivial pole as $\omega \rightarrow 0, k \rightarrow k_F$ only when $\nu$ vanishes in the large $d$ limit. For $O(1)$ values of $\nu$ we showed that the full UV two-point function takes the rather uninteresting form (\ref{nu1match}). 
\begin{figure}
    \centering
    \includegraphics[scale=0.45]{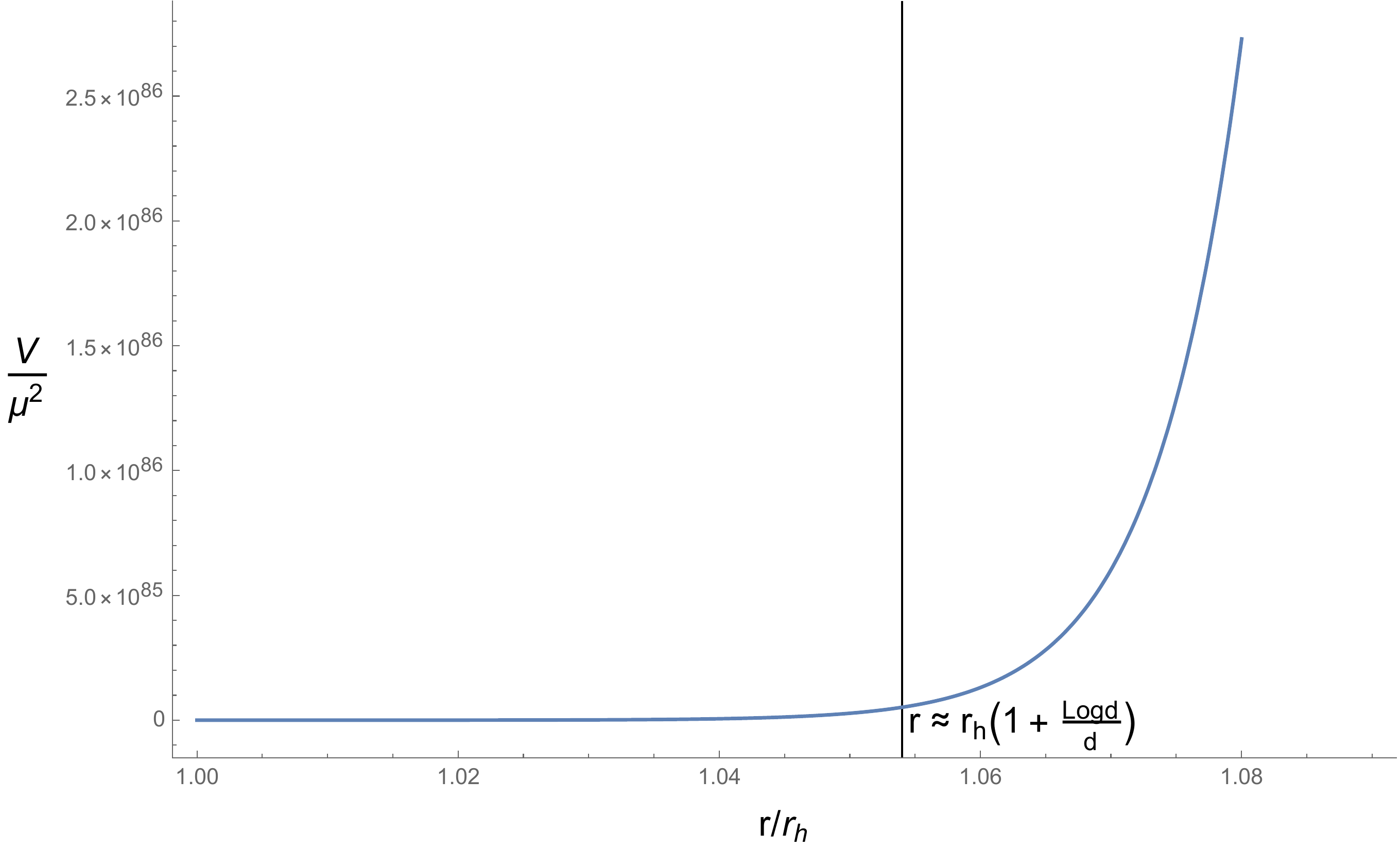}
    \caption{The potential for the Schrodinger problem for the bulk equation of motion for the scalar field $\phi(r,\omega,k)$ when $\nu \sim 1$. The potential monotonically increases and a mode that is (non) normalizable stays (non) normalizable throughout the geometry. }
    \label{nu1fig}
\end{figure}

The exponent $\nu$ can take $O(1)$ values in several cases. In the main text we considered the case when $m,q,k \sim O(1)$, in this appendix we will work with the case when $m,q$ and $k$ take generic $O(d)$ values. We will find that the result is similar to (\ref{nu1match}). The far-away solution (\ref{farphi}) has a similar expansion as in (\ref{farnear}). It is given by,

{\small
\begin{align}
    \phi\left(r_h\left(1 + \frac{ \log \rho}{d} \right),\omega,k\right)   = &  \frac{ N  }{ \left(2\pi d \nu r_h^d \rho \right)^{\frac{1}{2}} } \left( \frac{c_1(\omega,k) \cos\left( \theta   \right)+c_2(\omega,k) e^{i2 \theta }}{ \rho^{ \nu}  e^{\alpha } }  - 2 \rho^{ \nu}   e^{\alpha }  c_1(\omega,k) \sin\left( \theta \right)  \right), \label{farnear1}
\end{align}}
where,
\begin{align}
    N & \equiv 2^{-\frac{3 d}{4}} i^{- \sqrt{4 m^2+d^2}}  \left(q^2 r_h^2 - 2 k^2 + \frac{2 \sqrt{2} q r_h \omega}{d} + \frac{2 \omega ^2}{d^2}\right)^{d/4}, \quad \theta \equiv \frac{\pi d}{2} \sqrt{4 M^2 +1}, \nonumber \\
    \alpha & \equiv  d\sqrt{\frac{Q^2}{2}-\frac{ K^2}{r_h^2}+\nu^2  }  \log \left(\frac{\nu   +  \sqrt{\frac{Q^2}{2}- \frac{K^2}{r_h^2}+\nu^2  }}{\sqrt{\frac{Q^2}{2}-\frac{K^2}{r_h^2}}}\right) -  \nu  \left(\frac{\sqrt{2} Q r_h \omega }{Q^2 r_h^2 -2 K^2}+1\right).
\end{align}
We have used the notation $m\equiv M d, q  \equiv Q d $ and $k \equiv K d$. Meanwhile the near-horizon solution can be expanded near the $AdS_2$ boundary to obtain,
\begin{align}
    \phi(z,\omega,k) = b_1(\omega,k) (-z_h)^\frac{i z_h \omega}{r_h} \left( z^{\frac{1}{2}-\nu} -  \frac{z^{\frac{1}{2} +\nu }}{z_h^{2\nu}} G_{IR}(\omega,k)\right).  \label{nearfar1}
\end{align}
Matching the two we obtain,
{\scriptsize
\begin{align}
    \frac{c_2(\omega,k)}{c_1(\omega,k)} =  \left(d \sqrt{ M^2+\frac 1 4}\right)^{-d \sqrt{4 M^2+1}} e^{-i \pi  d \sqrt{4 M^2+1}} \left( G_{IR}(\omega,k) e^{-2 d \nu } \left(\frac{\sqrt{ M^2+ \frac 1 4} + \nu}{\sqrt{ M^2+\frac 1 4 - \nu ^2}}\right)^{d \sqrt{4 M^2+1}} \left(\frac{\kappa}{d}\right)^{2 \nu }- \frac{\cot \theta }{2} \right). \label{comp}
\end{align}}
For generic values of $m,q$ and $k$, the second term in the parenthesis is exponentially suppressed compared to the first. Using (\ref{farsol}), we thus obtain the UV two-point function, up to exponentially small corrections,
\begin{align}
   G_{UV} (\omega,k) =       G_{IR}(\omega,k)  \left(\frac{\sqrt{ M^2+ \frac 1 4} + \nu}{4r_h e^{ i \pi} \sqrt{ M^2+\frac 1 4 }}\right)^{2d \sqrt{ M^2+ \frac 1 4}} \left(\frac{\kappa}{d}\right)^{2 \nu } e^{-2 d \nu }. \label{unint}
\end{align}
We find that for $\nu \sim 1$, and generic $m,q,k \sim O(d)$ values, we obtain a similar looking Green's function as found before in (\ref{nu1match}). Thus this parameter regime is uninteresting from the perspective of finding Fermi surfaces. Figure \ref{nu1fig} shows the Schrodinger potential for the scalar field $\phi(r,\omega,k)$ when $\nu \sim 1$. The potential increases monotonically, and due to the large $d$ nature of our geometry, is of order $O(d^2)$ as soon as $\frac{r-r_h}{r_h} \gtrsim 1$. As a result a mode that is (non) normalizable in the near-horizon geometry stays (non) normalizable in the rest of the geometry. This is why the coefficients $b_1^+$ and $b_2^-$ are exponentially small. Subsequently the Green's function takes a form given in (\ref{unint}) and not the Fermi surface type given in (\ref{mit}).

Note that the terms in the parentheses in (\ref{comp}) are of similar magnitudes, in large $d$ counting, when $\nu \sim 1/d^{1/3}$. This is precisely when we find Fermi surfaces.

%%%%%%%%%%%%%%%%%%%%%%%%%%%%%%%%%%%%%%%%%%%%%%%%%%%%%%%%%%%%%%%%%%%%%%%%%%%%%%%%%%%%
%%%%%%%%%%%%%%%%%%%%%%%%%%%%%%%%%%%%%%%%%%%%%%%%%%%%%%%%%%%%%%%%%%%%%%%%%%%%%%%%%%%%
%%%%%%%%%%%%%%%%%%%%%%%%%%%%%%%%%%%%%%%%%%%%%%%%%%%%%%%%%%%%%%%%%%%%%%%%%%%%%%%%%%%%
%%%%%%%%%%%%%%%%%%%%%%%%%%%%%%%%%%%%%%%%%%%%%%%%%%%%%%%%%%%%%%%%%%%%%%%%%%%%%%%%%%%%
%%%%%%%%%%%%%%%%%%%%%%%%%%%%%%%%%%%%%%%%%%%%%%%%%%%%%%%%%%%%%%%%%%%%%%%%%%%%%%%%%%%%
%%%%%%%%%%%%%%%%%%%%%%%%%%%%%%%%%%%%%%%%%%%%%%%%%%%%%%%%%%%%%%%%%%%%%%%%%%%%%%%%%%%%

\section{Fermions} \label{fermat}
In this section, we will deal with bulk fermions. As mentioned in the introduction, the physics is qualitatively similar to the case of scalars, but the details are slightly different. We begin with the bulk action of the fermions minimally coupled to gravity  \cite{Iqbal:2009fd,Hartnoll:2016apf},
\begin{align}
S_{fer} = - i \int d^{d+1}x \sqrt{-g} \left( \bar\Psi \Gamma^\mu \left( \partial_\mu + \frac 1 4 \omega_{\mu ab} \Gamma^{ab}  - i q A_\mu \right)\Psi - m \bar \Psi \Psi  \right)
\end{align}
where,
\begin{align}
\Gamma^{ab}=\frac 1 2 \left[ \Gamma^a,\Gamma^b \right], \qquad \omega_\mu^{ab} = e_\nu^a \Gamma^{\nu}_{\sigma \mu} e^{\sigma b} - e^{\nu b} \partial_\mu e_\nu^a.
\end{align}
The $\Gamma^\mu$ are the $d+1$ dimensional gamma matrices, $\omega_\mu^{ab}$ is the spin connection and $\Gamma^\mu_{\sigma \rho}$ are the Christoffel symbols. For a diagonal metric that only depends on the bulk radial coordinate $r$, we can make the following redefinition,
\begin{align}
\Psi(r,t,x^i) = (-g g^{rr})^{- \frac 1 4} \psi(r,t,x^i). \label{refer}
\end{align}
This cancels the spin connection term in the bulk equation of motion, and we get,
\begin{align}
e_a^\mu \Gamma^a  \left( \partial_\mu  - i q A_\mu\right)\psi(r,t,x^i) = m \psi(r,t,x^i). \label{dirac}
\end{align}

We work with non-relativistic fermions that explicitly break Lorentz invariance as mentioned near footnote \ref{nonrel}. %we consider bulk fermions that live in $d+1$ spacetime dimensions but transform in the spin-1/2 representation of $SU(2)$.
Setting the vielbein to $e_a^\mu \equiv \sqrt{\left | g^{\mu\mu} \right|}$, and going to fourier space in the boundary coordinates, 
\begin{align}
    \psi(r,t,x^i) = \int d^{d-1}k \, d\omega \, e^{- i \omega t + i k_i x^i} \psi(r,\omega,k)
\end{align}
we obtain the bulk equation of motion for fermions in the black hole background (\ref{RN}),
\begin{align}
    \left(\partial_r  + i \sqrt{\frac{g^{tt}}{g^{rr}}} \sigma^2 \left( \omega + q A_t \right) \right) \psi(r,\omega,k) & = \left( \sqrt{\frac{g^{xx}}{g^{rr}}} \sigma^1 k_x \psi + \frac{m}{\sqrt{g^{rr}}} \sigma^3 \right) \psi. \label{fereom}
\end{align}
The $\sigma_i$ are the Pauli matrices,
\begin{align}
\Gamma^z=\sigma^3= \begin{pmatrix} 
1 & 0 \\	
0 & -1 
\end{pmatrix}  , \quad \Gamma^t=i\sigma^1=\begin{pmatrix} 
0 & i \\	
i & 0 
\end{pmatrix}, \quad \Gamma^x=-\sigma^2= \begin{pmatrix} 
0 & i \\	
-i & 0 
\end{pmatrix},
\end{align}
and $\psi(r,\omega,k)$ is a two component Weyl spinor. We will now solve this equation in the near-horizon region and the far-away region, and match them in the middle to obtain the full two-point function.

\subsection{Near-horizon region}
The near-horizon metric is given by (\ref{nh1}), that we reproduce here for the reader's convenience,
\begin{align}
     ds^2 = \frac{1}{d^2} \left( - \frac{\left(1- \frac{z}{z_h} \right)}{z^2} \frac{r_h^2}{\ell^2} d\tau^2 + \frac{\ell^2 dz^2}{z^2\left(1- \frac{z}{z_h}\right)}    \right)  + \frac{r_h^2}{\ell^2} dx_{d-1}^2.
\end{align}
The equation of motion (\ref{fereom}), can be exactly solved in the near-horizon geometry \cite{Faulkner:2009wj}. Throughout this appendix we will solve for the top component of $\psi(r,\omega,k^i)$. The top component satisfies a second order differential equation, the solution to which is given by,

{\footnotesize
\begin{align}
    \psi(r,\omega,k^i) = & z^{\nu } \left(-\frac{z}{z_h}\right)^{-2 \nu } (z-z_h)^{\frac{i \omega }{4 \pi  T}} \left(b_1(\omega,k) \left(-\frac{z}{z_h}\right)^{2 \nu } \, _2F_1\left(\nu -\frac{i q}{\sqrt{2} d},\frac{i q}{\sqrt{2} d}+\nu +\frac{i \omega }{2 \pi  T}+\frac{1}{2};2 \nu +1;\frac{z}{z_h}\right) \right. \nonumber \\
    & \left. +b_2(\omega,k) \, _2F_1\left(-\frac{i q}{\sqrt{2} d}-\nu ,\frac{i q}{\sqrt{2} d}-\nu +\frac{i \omega }{2 \pi  T}+\frac{1}{2};1-2 \nu ;\frac{z}{z_h}\right)\right),
\end{align}}
where, 
\begin{align}
    \nu = \frac{1}{d} \sqrt{m^2 - \frac{q^2}{2} + \frac{k^2}{r_h^2} }.
\end{align}
At the event horizon, this solution has an incoming mode $(z-z_h)^{-\frac{i \omega  z_h \ell ^2}{d r_h}}$ and an outgoing mode $(z-z_h)^{\frac{i \omega  z_h \ell ^2}{d r_h}}$. Imposing incoming boundary conditions at the horizon we obtain,

{\scriptsize
\begin{align}
    \psi(r,\omega,k^i) = & b_1(\omega,k) \Gamma (1+2 \nu ) z^{-\nu } (z-z_h)^{\frac{i \omega }{4 \pi  T}} \left[ z^{2\nu} \, _2\tilde{F}_1\left(\nu -\frac{i q}{\sqrt{2} d},\frac 1 2 + \nu + \frac{i q}{\sqrt{2} d} +\frac{i \omega }{2 \pi  T};1+2 \nu;\frac{z}{z_h}\right) \right. \nonumber  \\
    & \left. -\frac{z_h^{2 \nu } \Gamma \left(\frac{i q}{\sqrt{2} d}-\nu +1\right) \Gamma \left(\frac 1 2 - \nu -\frac{i  q}{\sqrt 2 d}-\frac{i \omega }{2 \pi  T}\right)}{\Gamma \left( \frac{i q}{\sqrt{2} d}+\nu +1\right) \Gamma \left(\frac 1 2 +\nu -\frac{i q}{\sqrt{2} d}-\frac{i \omega }{2 \pi  T}\right)} \, _2\tilde{F}_1\left(-\frac{i q}{\sqrt{2} d}-\nu , \frac 1 2 - \nu +\frac{i q}{\sqrt 2 d } +\frac{i \omega }{2\pi  T};1-2 \nu ;\frac{z}{z_h}\right) \right], \label{nearfer}
\end{align}}
where $ {}_2\tilde F_1$ are the regularized hypergeometric functions. %we have used the relation (\ref{refer}), $\Psi = $.

Like the scalars, we can imagine a $CFT_1$ that lives at the asymptotic boundary of the near-horizon $AdS_2$ geometry. The two-point function in this $CFT_1$ which we refer as the IR Green's function, is then given by,\footnote{The two-point function for spinors is also given by the ratio of the normalizable mode to the non-normalizable mode. However, both components of the spinor contribute to the two-point function. We will only present the answer here, and refer the reader to the standard work \cite{Iqbal:2009fd} and references therein for the holographic prescription for computing the Green's functions of spinors.}

{\small
\begin{align}
    G_{IR}(\omega,k^i) = - T^{2 \nu } \frac{\Gamma (1-2 \nu )  \Gamma \left(\frac{i q}{\sqrt{2} d}+\nu +1\right) \Gamma \left(\frac 1 2 + \nu-\frac{i q}{\sqrt{2} d}-\frac{i  \omega }{2 \pi  T}\right)}{\Gamma (1+2 \nu ) \Gamma \left(\frac{i q}{\sqrt{2} d}-\nu +1\right) \Gamma \left(\frac 1 2 - \nu -\frac{i  q}{ \sqrt 2 d}-\frac{i  \omega }{2\pi  T}\right)} \frac{ \frac{q}{\sqrt 2 d} + i \nu - \frac{m}{d} + \frac{ i k}{d r_h}}{ \frac{q}{\sqrt 2 d} - i  \nu - \frac{m}{d}  +\frac{ i k}{d r_h }} \left( \frac{4\pi}{\kappa r_h}\right)^{2\nu}.
\end{align}}

\subsection{Far-away region}
The far-away region in the large $d$ limit, is extremely simple and we just obtain vacuum $AdS_{d+1}$ with a constant gauge potential,
\begin{align}
      ds^2   = - \frac{r^2}{\ell^2}  dt^2   + \frac{\ell^2}{r^2} dr^2 + \frac{r^2}{\ell^2} dx^2_{d-1}, \qquad A_t = \frac{r_h}{2\ell^2}.
\end{align}
The fermion equation of motion (for the top component) in this background is a standard differential equation. The solution is given in terms of Bessel functions,

{\footnotesize
\begin{align}
    \psi(r,\omega,k^i) = & \frac{\sqrt[4]{2 k^2-q^2 r_h^2-2 \sqrt{2} q r_h \omega -2 \omega ^2}}{2^{3/4} i^{m} \sqrt{r}} \left[c_1(\omega,k) \Gamma \left(\frac{1}{2}-m\right) I_{-m-\frac{1}{2}}\left(\frac{i \sqrt{\frac{q^2 r_h^2}{2}-k^2+\sqrt{2} q r_h \omega +\omega ^2}}{r}\right) \right. \nonumber \\
    & \left. + c_2(\omega,k) i^{2 m+1} \Gamma \left(m+\frac{3}{2}\right) I_{m+\frac{1}{2}} \left(\frac{i \sqrt{\frac{q^2 r_h^2}{2}-k^2+\sqrt{2} q r_h \omega +\omega ^2}}{r}\right)\right]. \label{farfer}
\end{align}}
Using the asymptotic expansion of the Bessel function, and recalling our redefinition (\ref{refer}), we have at the asymptotic $AdS_{d+1}$ boundary,
\begin{align}
    \lim_{r \rightarrow \infty}\Psi(r,\omega,k^i) = & c_1(\omega,k) \left(\frac{i}{2}\right)^{-m} r^{-\frac{d}{2}+m} \left(k^2-\frac{q^2 r_h^2}{2}-\sqrt{2} q r_h \omega -\omega ^2\right)^{-\frac{m}{2}} \nonumber \\
    & + \frac{c_2(\omega,k)  i^{m+1}}{2^{m+1}} r^{-\frac{d}{2}-m-1}  \left(k^2-\frac{q^2 r_h^2}{2}-\sqrt{2} q r_h \omega -\omega ^2\right)^{ \frac{m+1}{2}} .
\end{align}
Taking the ratio of the normalizable mode to the non-normalizable mode (while including both components of the spinor) we obtain the fermion two-point function in the $CFT_d$,
\begin{align}
    G_{UV} (\omega,k) %& = i   2^{-3 m-\frac{3}{2}} e^{i \pi  m} \left(2 k^2-q^2 r_h^2-2 \sqrt{2} q r_h \omega -2 \omega ^2\right)^{m+\frac{1}{2}} \frac{c_2(\omega,k)}{c_1(\omega,k)}. \nonumber \\
    & = \frac{i  e^{i \pi  m} (2 m+1) \left( k^2- \frac{q^2 r_h^2}{2}-  \sqrt{2} q r_h \omega - \omega ^2\right)^{m+\frac{1}{2}}}{  2^{2 m } \left(-2 k+\sqrt{2} q r_h+2 \omega \right)} \frac{c_2(\omega,k)}{c_1(\omega,k)}. \label{uvfer}
\end{align}
To obtain the full UV two-point function we need to obtain the integration constants $c_i(\omega,k)$. We will obtain them by matching the far-away solution (\ref{farfer}) with the near-horizon solution (\ref{nearfer}) in the following sections, just like we did for the scalars.

\subsection{Parametric regime of interest} \label{param}
Before we proceed with the matching calculation, we would like to reiterate the relevant parametric regimes of interest as discussed near (\ref{nusc}). Recall that the scaling exponent of the fermions is given by,
\begin{align}
    \nu = \frac{1}{d} \sqrt{m^2 - \frac{q^2}{2} + \frac{k^2}{r_h^2}},
\end{align}
while the dimension of the fermionic operator in the $CFT_d$ is given by,
\begin{align}
    \Delta = \frac d 2 + m \ell .
\end{align}

The parameters $m,q$ and $k$ can have any dependence on $d$, since that is not fixed by the symmetries or the dynamics of the theory. We will restrict to the cases that we find are the most interesting. In a $d$ dimensional theory, we expect the momentum $k\equiv |k^\mu|$ to scale as $\sqrt{d}$. %This is obvious if we assume that the individual momentum components $k^i$ are $O(1)$ in large $d$ scaling. 
As discussed in the main text, this regime also makes sense from the perspective of the DMFT ansatz, where the hopping parameter is rescaled as $t_{ij} \rightarrow t_{ij}/\sqrt{d}$. We find from the matching calculation that if $k\sim \sqrt{d}$, and if we want the UV two-point function to take the form of (\ref{mit}), then $m$ and $q$ also need to scale as $\sqrt{d}$.  In such a parametric regime, we find bulk Fermi surfaces when,
\begin{align}
    \nu = \frac{V(m,q,k_F)}{d^{2/3}}, \label{fernu}
\end{align}
where $V(m,q,k_F)$ is an $O(1)$ number, with at most logarithmic dependence on $d$. In our opinion, this is the most natural parametric regime for the fermions.

If we had instead taken $m,q,k \sim 1$, we would have found bulk Fermi surfaces for $\nu \sim V(m,q,k_F)/d$, where $V(m,q,k_F)$ is some $O(1)$ number. This is also an interesting case to consider, but we shall not pursue it here since it gives rise to the same kind of physics. We could have also considered the scenario where $m,q,k$ scale with $d$, like we did for the scalars. In that case we obtain physics similar to the scalars, and find that $\nu = V(m,q,k_F)/{d^{1/3}}$, when looking for Fermi surfaces. Finally, just like the scalars, the fermion two-point function takes the uninteresting form (\ref{nu1match}), and not the Fermi surface type (\ref{mit}), whenever $\nu \sim 1$. %This is the largest value of $\nu$ that we can obtain for the range of $\Delta$ that we are interested in. Through out this paper we have restricted ourselves to the case where $\Delta \sim O(d)$ or smaller, since that corresponds to marginal or relevant operators in the $CFT_d$. In the rest of the appendix we will perform the matching calculation for fermions when $m,q,k \sim \sqrt{d}$ and obtain (\ref{fernu}).

%$m,q,k$ do not scale with $d$, the second where $m,q,k$ grow with $d$ and the th 

\subsection{Finding bulk Fermi surfaces}
Just as we did for the scalars, we will rewrite our far-away solution and the near-horizon solution using the coordinates for the middle region $(\rho,t)$. The radial coordinate of the far-away region is related to the mid-region coordinate by,
\begin{align}
    \frac{r}{r_h} =  \rho^{\frac{1}{d-2}}.
\end{align}
We want to expand the far-away solution in the middle region. This corresponds to taking $r \rightarrow r_h$. In terms of the middle region coordinates this means,
\begin{align}
    r = r_h \left( 1+ \frac{\log \rho}{d} \right).
\end{align}
The far-away solution is valid as long as,
\begin{align}
    \rho \gtrsim d. \label{farvalf}
\end{align}
Meanwhile the near-horizon coordinate $z$ is given by,
\begin{align}
    \rho =  \frac{\uu}{d} \left( \frac{z}{z_h} - 1\right) + 1. 
\end{align}
The near-horizon is valid as long as,
\begin{align}
    \rho - 1 \ll 1 \label{nearvalf}
\end{align}
Clearly the regimes of validity (\ref{farvalf}) and (\ref{nearvalf}) have no overlap. 

There are perhaps multiple ways of proceeding with this problem. We will take the simplest approach in our opinion. We will solve the bulk equation of motion in the middle region perturbatively in $1/d$ and match it with the near and far solutions. At each order we get two new integration constants and two boundary conditions, thus the mid-region solution is uniquely fixed. The metric in the middle region is given by (\ref{drn}),
\begin{align}
    ds^2= -\frac{(\rho - u^2)(\rho-1)}{\rho^2} \frac{r_h^2}{\ell^2} dt^2 + \frac{\ell^2 d\rho^2}{d^2(\rho - u^2)(\rho-1)}   +  \frac{r_h^2}{\ell^2} dx_{d-1}^2. 
\end{align}
Let the bulk fermion field be given by the following perturbative expansion,
\begin{align}
    \psi(\rho,\omega,k) = \chi(\rho,\omega,k) + \frac{1}{d^{1/6}} \chi_1(\rho,\omega,k) + \frac{1}{d^{1/3}} \chi_2(\rho,\omega,k) + \frac{1}{d^{1/2}} \chi_3(\rho,\omega,k) + \ldots 
\end{align}
The bulk equation of motion can now be solved order by order in $1/d$. Each $\chi_i$ satisfies a second order differential equation. It also needs to satisfy two matching conditions, one with the near-horizon solution and the other with the far-away solution. This uniquely fixes each $\chi_i(\rho,\omega,k)$, and gives us $c_i(\omega,k)$ in terms of $b_1(\omega,k)$ and $G_{IR}(\omega,k)$. Performing this calculation we find the coefficients $c_i(\omega,k)$ that are needed to obtain the full UV two-point function (\ref{uvfer}). We will not write down the rather long expressions for $c_i(\omega,k)$ for arbitrary values of $k$ but only write down the final answer near the near Fermi surface $k=k_F$. As argued near (\ref{bulkg}), the ratio $c_i(\omega,k)$ takes the form,
\begin{align}
    \frac{c_2(\omega,k)}{c_1(\omega,k)} & = \frac{b_2^-(0,k_F)}{\omega + v_F(k-k_F) + b_1^+ (0,k_F) G_{IR} (\omega,k_F)}. 
\end{align}
The constants $b_i$ are given by,

{\footnotesize
\begin{align}
    b_2^- & = \frac{M r_h \tan \left(\pi  \sqrt{d} M\right) \left(\text{Bi}(x) \left(33600 M^2 \sqrt{x} \log \left(\frac{d}{\kappa ^2}\right)+5376 x^5+12535 x^2\right)+10 \left(1416 x^3+223\right) \text{Bi}'(x)\right)}{140 \sqrt{2} i^{4M\sqrt d} Q \left(137 x^2 \text{Ai}(x)+2 \left(24 x^3+25\right) \text{Ai}'(x)\right)}, \nonumber \\
    %\frac{r_h\tan \left(\pi  \sqrt{d} M\right) \left(V \text{Bi}\left(\frac{V^2}{2^{2/3} M^{4/3}}\right) \left(67200 M^8 \log \left(\frac{d}{\kappa ^2}\right)+12535 M^4 V^3+1344 V^9\right)+20 \sqrt[3]{2} M^{8/3} \left(223 M^4+354 V^6\right) \text{Bi}'\left(\frac{V^2}{2^{2/3} M^{4/3}}\right)\right)}{70\ 2^{5/6} M^{5/3} i^{4 m} Q \left(137\ 2^{2/3} M^{4/3} V^4 \text{Ai}\left(\frac{V^2}{2^{2/3} M^{4/3}}\right)+8 \left(25 M^4+6 V^6\right) \text{Ai}'\left(\frac{V^2}{2^{2/3} M^{4/3}}\right)\right)} \nonumber \\ 
    b_1^+ & = \frac{960 \sqrt[6]{2} M^{19/3} r_h V \text{Ai}\left(x\right) \log \left(\frac{d}{\kappa^2}\right)}{Q \left(137\ 2^{2/3} M^{4/3} V^4 \text{Ai}\left(x\right)+8 \left(25 M^4+6 V^6\right) \text{Ai}'\left(x\right)\right)}, \qquad x \equiv \frac{V^2}{2^{2/3}M^{4/3}}. \nonumber 
\end{align}}

The Fermi velocity and momentum is given by,
\begin{align}
    v_F = \frac{\sqrt{q^2-2 m^2}}{q}, \qquad  k_F = r_h \sqrt{\frac{q^2}{2}-m^2}.
\end{align}
Note that $q,m$ scale as $O(\sqrt d)$, thus the Fermi velocity is $O(1)$ while the Fermi momentum is $O(\sqrt{d})$. The scaling exponent is given by,
\begin{align}
    \nu = \frac{V(m,q,k_F)}{d^{2/3}},
\end{align}
where $V(m,q,k_F)$ satisfies the relation,
\begin{align}
    \frac{\text{Ai}'\left(\frac{V^2}{2^{2/3} M^{4/3}}\right)}{\text{Ai}\left(\frac{V^2}{2^{2/3} M^{4/3}}\right)} = - \frac{V \left(67200 M^8 \log \left(\frac{d}{\kappa^2} \right)+12535 M^4 V^3+1344 V^9\right)}{20 \sqrt[3]{2} M^{8/3} \left(223 M^4+354 V^6\right)}. \label{ferai}
\end{align}

\section{Asymptotic expansion of Bessel functions} \label{bessec}

We will be using the following identities quite frequently,
\begin{align}
    %I_{-\nu}(z) & = I_{\nu}(z)+ \frac{2}{\pi} \sin (\nu \pi) K_{\nu}(z),   \nonumber \\
    I_{\nu}(z) & = e^{\mp \frac{\nu \pi i}{2}} J_\nu \left( z e^{i \frac{\pi}{2}} \right),  \qquad 
    %K_\nu(z) & = \pm \frac{1}{2} \pi i e^{\pm \frac{\nu \pi i }{2}} H^{\pm}_{\nu}(z e^{\pm \frac{\pi i }{2} }) \nonumber \\
    J_{\nu} (z e^{m \pi i})  = e^{m \nu \pi i } J_\nu(z), \nonumber \\
    Y_{\nu} (z e^{m \pi i}) & = e^{- m \nu \pi i} Y_{\nu} (z) + 2 i \sin(m \nu \pi) \cot(\nu \pi) J_\nu(z). \label{signbes}
\end{align}
In particular, we will be using the expansion of Watson for asymptotically large argument and large order i.e. $\nu,z \rightarrow \infty$ with $\nu/z$ fixed,
\begin{align}
    J_\nu(z) & = \frac{\exp \left(\sqrt{\nu ^2-z^2}-\nu  \text{sech}^{-1}\left(\frac{z}{\nu }\right)\right)}{\sqrt{2 \pi  \sqrt{\nu ^2-z^2}}} \sum_{m=0}^\infty \frac{\Gamma \left(m+\frac{1}{2}\right) A_m}{\Gamma \left(\frac{1}{2}\right) \left(\frac{1}{2} \sqrt{\nu ^2-z^2}\right)^m}, \\
    Y_\nu(z) & = \frac{\exp \left(\nu  \text{sech}^{-1}\left(\frac{z}{\nu }\right)-\sqrt{\nu ^2-z^2}\right)}{\sqrt{2 \pi  \sqrt{\nu ^2-z^2}}} \sum _{m=0}^\infty (-1)^{m+1} \frac{\Gamma \left(m+\frac{1}{2}\right)  A_m}{\Gamma \left(\frac{1}{2}\right) \left(\frac{1}{2} \sqrt{\nu ^2-z^2}\right)^m}. \label{besexp}
\end{align}
where,
\begin{align}
    A_0 & = 1, \quad A_1 = \frac{1}{8}-\frac{5}{24} \left(\frac{\nu }{\sqrt{\nu ^2-z^2}}\right)^2, \\
    A_2 & =\frac{3}{128} -\frac{77}{576} \left(\frac{\nu }{\sqrt{\nu ^2-z^2}}\right)^2  +\frac{385}{3456}\left(\frac{\nu }{\sqrt{\nu ^2-z^2}}\right)^4.
\end{align}
When we tune near the Fermi surface we will need the formula for asymptotic expansions at large order $\nu$ and large argument $z$ such that $|\nu - z | \sim \nu^{1/3}$. This requires a different expansion than Watson's and is given by the result of Olver \cite{olver_1952},  

{\small
\begin{align}
    J_\nu(\nu + \tau^{\frac 1 3} \nu) =  \frac{2^{\frac 2 3}  \text{Ai}'\left(-2^{\frac 1 3} \tau\right)}{\nu^{\frac 1 3}  } \left( \frac{3 \tau^2}{10 \nu^{2/3}} + O\left( \frac{1}{\nu^{4/3}}\right) \right) +\frac{ 2^\frac{1}{3} \text{Ai}\left(-2^{\frac 1 3} \tau\right)}{\nu^{\frac 1 3}}  \left(1-\frac{\tau}{5 v^{2/3}} + O\left( \frac{1}{\nu^{4/3}}\right)  \right), \nonumber \\ 
    Y_\nu(\nu + \tau^{\frac 1 3} \nu) = -\frac{2^{\frac 2 3} \text{Bi}'\left(-2^{\frac 1 3} \tau\right)}{ \nu^{\frac 1 3}} \left( \frac{3 \tau^2}{10 \nu^{2/3}} + O\left( \frac{1}{\nu^{4/3}}\right) \right)  -\frac{ 2^\frac{1}{3} \text{Bi}\left(-2^{\frac 1 3} \tau\right)}{\nu^{\frac 1 3}}  \left(1-\frac{\tau}{5 v^{2/3}} + O\left( \frac{1}{\nu^{4/3}}\right)  \right), \label{hailolver}
\end{align}}
where $\text{Ai}(z)$ and $\text{Bi}(z)$ are the Airy functions of the first and second kind respectively.

\bibliographystyle{JHEP}
\bibliography{bibli}

\end{document}